\documentclass[twocolumn,epjc3]{svjour3}

\usepackage{epsfig}
\usepackage{lineno}
\usepackage{amsmath}
\usepackage{todonotes}
\RequirePackage[T1]{fontenc}

\smartqed  

\RequirePackage{graphicx}
\RequirePackage{mathptmx}      
\RequirePackage{flushend}
\RequirePackage[numbers,sort&compress]{natbib}
\RequirePackage[colorlinks,citecolor=blue,urlcolor=blue,linkcolor=blue]{hyperref}

\journalname{Eur. Phys. J. C}
\begin{document}


\title{Cryogenic scintillation veto for $0\nu2\beta$ bolometric experiments: from a concept to BINGO prototypes}

\author{A.~Armatol\thanksref{addr1,e1}
        \and
        D.~Baudin\thanksref{addr1}
        \and
        G.~Benato\thanksref{addr4}
         \and
        V.~Berest\thanksref{addr1,e2}
         \and
         M.~Buchynska\thanksref{addr5}
         \and
         P.~Carniti\thanksref{addr8}
         \and
F.A.~Danevich\thanksref{addr9,addr10,addr14} 
        \and 
        M.~De~Jesus\thanksref{addr2}
        \and 
         F.~Ferri\thanksref{addr1}
        \and 
         A.~Giuliani\thanksref{addr5}
        \and 
         H.~Gomez\thanksref{addr1}
        \and 
         C.~Gotti\thanksref{addr8}
        \and 
         Ph.~Gras\thanksref{addr1}
        \and 
         H.~Khalife\thanksref{addr1,e3}
        \and 
        V.V.~Kobychev\thanksref{addr9}
        \and
        D.S.~Kovalenko\thanksref{addr9}
        \and 
        M.~Lefevre\thanksref{addr1}
        \and
        Ph.~Mas\thanksref{addr1}
        \and
        E.~Mazzucato\thanksref{addr1}
        \and
        J.F.~Millot\thanksref{addr1}
        \and
         C.~Nones\thanksref{addr1}
         \and        
         G.~Pessina\thanksref{addr8}
        \and          
         D.V.~Poda\thanksref{addr5} 
        \and          
        B.~Schmidt\thanksref{addr1}  
        \and
        O.~Tellier\thanksref{addr1}
        \and 
        V.I.~Tretyak\thanksref{addr4,addr9,addr10}
        \and
        D.~Vurpillot\thanksref{addr1}
        \and
         M.M.~Zarytskyy\thanksref{addr9}
        \and
         A.~Zolotarova\thanksref{addr1}
        }

\thankstext{e1}{Now at: Institut de Physique des 2 Infinis, Villeurbanne, France}
\thankstext{e2}{Now at: University of California, Berkeley, CA, USA}
\thankstext{e3}{Now at: IJCLab, CNRS/IN2P3, Orsay, France}

\institute{IRFU, CEA, Universit\'{e} Paris-Saclay, F-91191 Gif-sur-Yvette, France \label{addr1}
\and
     INFN Laboratori Nazionali del Gran Sasso, I-67100 Assergi (AQ), Italy \label{addr4}
\and
     Universit\'{e} Paris-Saclay, CNRS/IN2P3, IJCLab, 91405 Orsay, France \label{addr5}
\and
     INFN Sezione di Milano-Bicocca, I-20126 Milano, Italy \label{addr8}
\and
     Institute for Nuclear Research of NASU, 03028 Kyiv, Ukraine \label{addr9}
\and     
    Institute of Experimental and Applied Physics, Czech Technical University in Prague, 110 00 Prague 1, Czech Republic \label{addr10}
\and
    Gran Sasso Science Institute, I-67100 L’Aquila, Italy \label{addr14}
\and
    Univ Lyon, Universit{\'e} Lyon 1, CNRS/IN2P3, IP2I-Lyon, 69100 Villeurbanne, France \label{addr2}
}

\date{Received: date / Accepted: date}

\maketitle

\begin{abstract}
Achieving a very low, ideally zero, background level, along with ensuring high energy resolution and detection efficiency, are the main tasks in the development of the next generation of experiments to search for neutrinoless double beta ($0\nu2\beta$) decay. Cryogenic scintillating bolometers meet these basic requirements perfectly. The main ideas pursued in the BINGO (Bi-Isotope $0\nu2\beta$ Next Generation Observatory) project are the use of a cryogenic scintillation anti-coincidence veto, the minimization of the detector array passive materials, and the use of advanced cryogenic light detectors to reduce backgrounds from surface radioactivity and random signal coincidences.
This article presents simulation results for detectors based on Li$_2$MoO$_4$ or TeO$_2$ surrounded by a 5-cm-thick active Bi$_4$Ge$_3$O$_{12}$ scintillation veto, demonstrating the potential to reduce the background by up to a factor of $10^3$, depending on the source location and the type of radionuclide. We further present results from the first cryogenic tests of a small scale prototype with respect to the feasibility of this proposed veto.
\end{abstract}

\section{Introduction} 

The discovery of neutrino mass in experiments sensitive to neutrino flavour oscillations (see details and references in e.g. \cite{King:2014,Bellini:2014,Balantekin:2013}) was the first clear evidence of an effect beyond the Standard Model of particles and interactions (SM) \cite{Bilenky:2015a,Miranda:2015}. Further complementary research on neutrinos may reveal additional possibilities for SM extensions. Neutrinos remain the  least understood fundamental particles, whose investigation can provide answers to several essential questions about the creation and structure of the Universe. First, experiments measuring neutrino oscillations cannot currently determine the absolute value of the neutrino mass, nor the hierarchy of neutrino mass eigenvalues. There are three ways to measure the mass of neutrino: inference of the neutrino mass from astrophysical and cosmological data \cite{Xu:2021,Shao:2025}, measurements of the neutrino mass in beta decay \cite{Formaggio:2021,Aker:2025} and electron capture \cite{Alpert:2025}, and the double beta decay \cite{Bilenky:2015b,Vergados:2016,Dolinski:2019,Gomez:2023,Agostini:2023,Bossio:2024,Giuliani:2026}. All these methods complement each other, as they are sensitive to different combinations of the neutrino mass eigenvalues. Cosmological estimates determine the sum of neutrino mass eigenvalues dependent on the cosmological model and data choice, measurements of the tritium beta decay or $^{163}$Ho electron capture spectra near their endpoint provide the incoherent sum of neutrino masses, while the neutrinoless double beta ($0\nu2\beta$) decay is sensitive to the effective Majorana mass of the neutrino that is a coherent sum of the neutrino mass eigenvalues ($\langle m_{\nu}\rangle$; in the Majorana scheme \cite{Majorana:1937} the neutrino and antineutrino are identical). 

Despite significant efforts since the late 1940s, the $0\nu2\beta$ decay has yet to be observed. The most sensitive experiments give only the half-life limits on several, most promising nuclides from the experimental and theoretical points of view, resulting in the constraints on the effective Majorana neutrino mass in the range $\langle m_{\nu}\rangle\sim$ (0.03--0.6) eV \cite{Adams:2025a,Azzolini:2022,Augier:2022,Abe:2025,Agrawal:2025a,Acharya:2026}, depending on the nuclide and the nuclear matrix elements used to estimate the $\langle m_{\nu}\rangle$.

The aim of the current $0\nu2\beta$ projects, which are in the preparation stage, is to achieve a sensitivity to the neutrino mass that would allow either to register the process in the case of the inverted scheme of the mass eigenvalues, $\langle m_{\nu}\rangle\sim$ (0.02--0.05) eV, or to reject this scheme in favour of the normal one,  $\langle m_{\nu}\rangle <0.02$ eV. The inverted scheme neutrino mass region corresponds to the half-lives of $T_{1/2}\sim10^{27}$--10$^{28}$ yr for the most promising nuclides, or several decays over several years of measurements in a $\sim1$-ton detector containing the isotope under study. One of the most promising approaches to measure such an extremely low activity is the technique of low-temperature scintillation bolometers, which are characterized by high energy resolution and detection efficiency thanks to the calorimetric approach, where the studied nuclides are part of the detector compound \cite{Alfonso:2025,Agrawal:2025b,Alfonso:2026}.  

Reducing the background, preferably to zero, is the most important and, at the same time, the most challenging task for any $0\nu2\beta$ experiment, including low-tempera\-ture ones. After all the efforts to carefully select the materials of the detector and passive screens in terms of their radiopurity and to place the experimental setup deep underground to protect against cosmic radiation, 
there is still remaining background. 
The primary sources of this background of low-temperature detectors are surface contamination of the crystals containing the isotope of interest and of surrounding materials, 
and the very low, but still significant, radioactive contamination of crystals and the passive elements such as the mechanical detector support, cryostat shields and the readout traces of the detector module. 
For experiments using $^{100}$Mo-enriched low-temperature detectors, random coincidences of the two-neutrino $2\beta$ decay of the nuclide under study can significantly contribute to the background too \cite{Chernyak:2012}. 
The CUPID and AMoRE-II projects, currently in the preparation stage, anticipated a background index of $\sim10^{-4}$ counts/(keV $\cdot$ kg $\cdot$ yr) in the region of interest (ROI) \cite{Alfonso:2025,Agrawal:2025b}.

The BINGO (Bi-Isotope $0\nu2\beta$ Next Generation Observatory) project \cite{BINGO-web,Khalife:2023} was proposed to further reduce the background by at least an order of magnitude. 
This should provide conditions for conducting a zero-background ton-scale experiment, which will enable the exploration of the normal scheme of neutrino mass eigenvalues, a concept intuitively perceived by many physicists as being realized in Nature. One of BINGO’s key ideas is to surround the low-temperature detector array with a scintillation veto cooled to milli-Kelvin temperature; this possibility is investigated in the present paper. 

Preliminary simulations evaluating the concept of the scintillation low-temperature veto to suppress the background of lithium molybdate (Li$_2$MoO$_4$) or tellurium dioxide (TeO$_2$)
detectors are given in Sect. \ref{concept}. The design of the BINGO prototype setup, and the first results of cryogenic scintillation veto tests are presented in Sect. \ref{results}. A summary and an outlook in Sect. \ref{conclusions} conclude this work.

\section{Cryogenic active veto concept}
\label{concept}

\subsection{Background reduction by scintillation veto}
\label{Idea-of-scint-active-veto}
 
Cryogenic detector arrays are designed to be surrounded with lead and copper screens. These screens are functioning as thermal and radiation shield between the detector arrays and cryogenics, experimental infrastructure, and the environment. These shields can be designed to suppress this environmental radiogenic background component to sub-dominant levels, but in turn intrinsic contaminations of the copper screens and lead shields become appreciable contributors to the background. The copper support structure of the detector array is another major contributor to the detectors' background.  A significant effort has been invested in the selection and manufacturing of ultra-pure materials \cite{Christofferson:2018}, but a further improvement faces significant technical challenges and is costly. 

Therefore, surrounding the cryogenic detector array with a layer of scintillators with a high $\gamma$-ray detection efficiency should suppress the background substantially. The active counter surrounding the detector assembly, working in anti-coincidence, will also allow the use of coincidence cuts between the detectors on the outside of the array and the cryogenic veto, thereby 
providing 
the suppression of the otherwise elevated background for these detectors and its supporting structure (mainly copper), especially from surface contamination of the materials. The latter is one of the most dangerous and difficult-to-reduce sources of background in low-temperature detectors.

\begin{figure}[!ht]
 \centering
 \mbox{\epsfig{figure=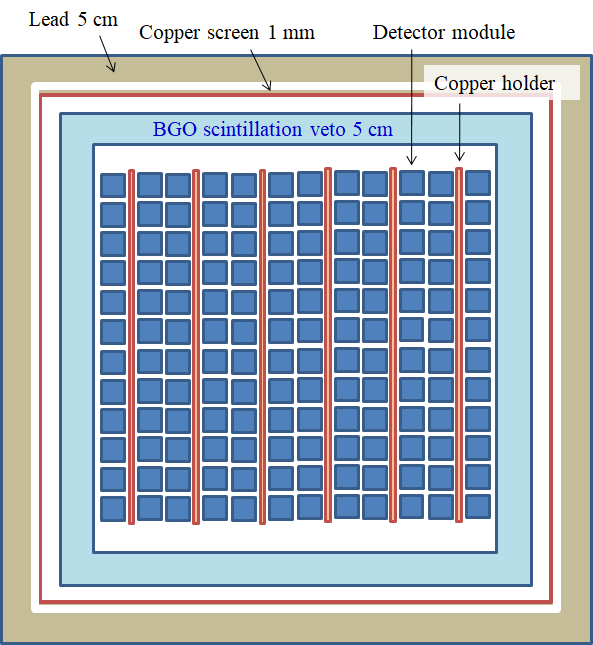,height=8.0cm}}
 \caption{A simplified scheme of a cryogenic experiment used for the Monte Carlo simulation of background levels from radioactive contamination of the setup and detector materials.}
\label{fig:setup-for-MC-sim}
\end{figure}

Monte Carlo simulations using the GEANT4 toolkit \cite{GEANT4} were performed to assess the potential of the active veto to reduce the background from radioactive contamination by thorium and radium for three representative volumes: the thermal screens, the inner lead layer around the copper screen, and copper support of the crystals detectors array. A simplified scheme of the setup used for the Monte Carlo simulation of background levels from radioactive contamination of the setup materials is shown in Fig. \ref{fig:setup-for-MC-sim}. A further study was made investigating the dependence of the effectiveness of the active veto on the scintillator material, and the energy threshold of the scintillation veto counter ($E_{\rm{th}}^{\rm{SC}}$).

The model of the experimental setup for simulations consisted of a 1-mm-thick copper screen surrounding a 5-cm-thick layer of the scintillation veto scintillators, 101.2 cm in diameter and 100.6 cm in height (outer dimensions). Bismuth germanium oxide (BGO, chemical formula Bi$_4$Ge$_3$O$_{12}$) crystals were chosen as the material for a scintillation veto (the choice of BGO will be justified in Sect. \ref{sec:sim-photons}). We assume that the scintillation crystals are not covered with any reflector. This makes it possible to detect $\alpha$ and $\beta$ particles and reduce the background caused by surface contamination of the crystals and of the supporting structure. The total mass of BGO crystals employed in this model is 1548 kg, and the mass of the copper screen is 44.6 kg. The detector array was modelled as a $12\times12\times12$ array of TeO$_2$ (in total 951 kg) or Li$_2$MoO$_4$ (in total 487 kg) 
crystals $4.5\times4.5\times4.5$ cm$^3$ with a 0.5-cm gap between the crystals, and between the crystals and the copper support. The sizes of the detector arrays are $59.5 \times 59.5\times 68.5$ cm$^3$. The larger size of the assembly in one direction occurs due to the presence of copper holders: the crystals are mounted in pairs on copper supports (two crystals on one support) with sizes $1.0\times 2.25\times 5.0$ cm$^3$. The total mass of the copper supports is 87.1 kg. For the simulations, we assumed a 5-cm-thick lead shield surrounding the copper screen with a 0.5-cm gap between the shield and the copper screen. The mass of the lead is 3158.8 kg. It should be noted that this setup geometry is somewhat simplified, but not too far from the design of BINGO \cite{Armatol:2024}, where the amount of materials to support the detectors is expected to be minimized as much as possible. The DECAY0 event generator \cite{DECAY0} was used to describe the kinematics of particles in decays of $^{208}$Tl and $^{214}$Bi 
--- the most dangerous natural radionuclides for $0\nu2\beta$ search experiments due to the large energy of the gamma rays emitted \cite{Gomez:2023} --- 
in the copper screen, and the copper supports of the detector array. 
Only decays of $^{208}$Tl were generated in the 
external 
lead layer. 

\begin{figure}[!ht]
 \centering
 \mbox{\epsfig{figure=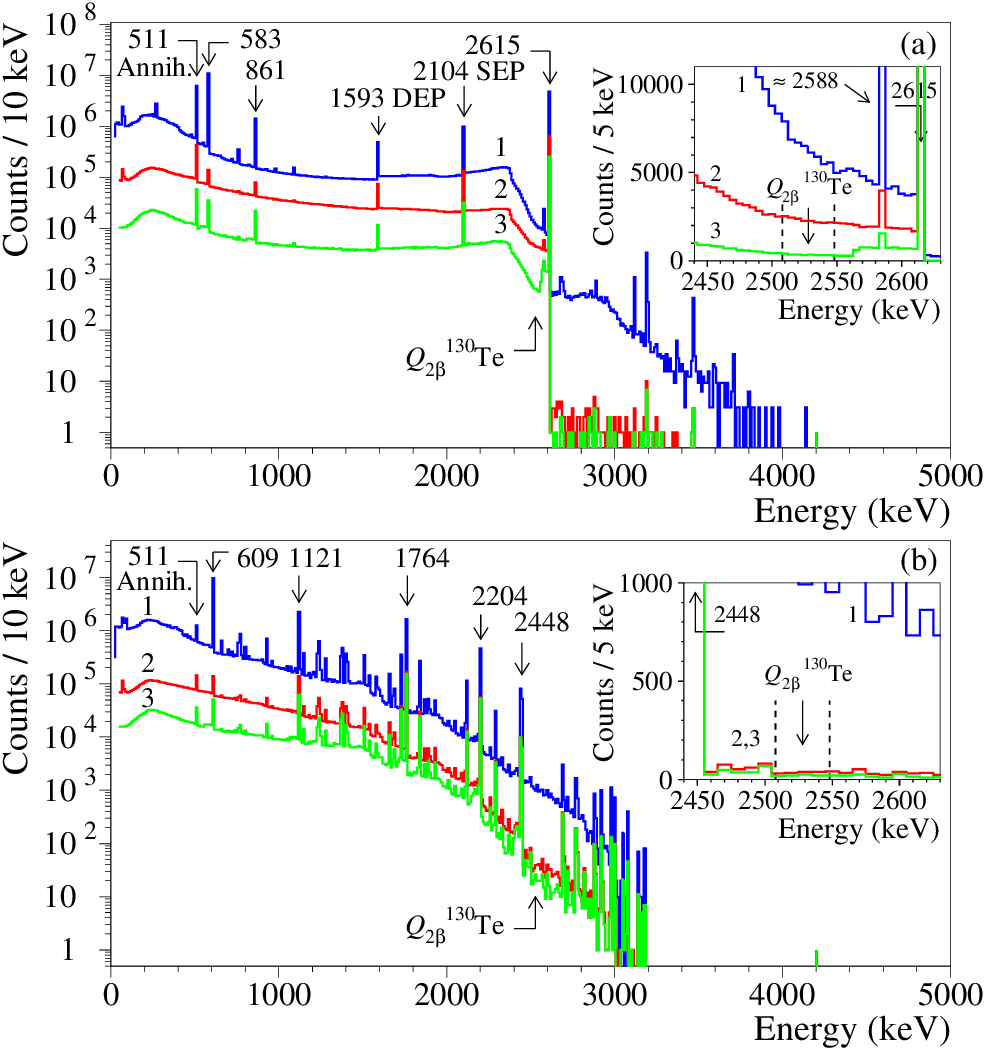,height=9.0cm}}
 \caption{Monte Carlo simulated energy spectra of the TeO$_2$ detector array for $1.6\times10^8$ $^{208}$Tl (a) and $2.6\times10^8$ $^{214}$Bi (b) decays in the copper screen: (1) without scintillation veto, (2) with BGO scintillators as passive material, (3) in anti-coincidence with the scintillation veto with a threshold of 50 keV. The most intense $^{208}$Tl and $^{214}$Bi peaks are labelled. SEP and DEP denote single and double escape peaks, respectively. Energy of the peaks is in keV. (a-Inset) The reduction of background with the active veto in the energy interval where a peak from $0\nu2\beta$ decay of $^{130}$Te is expected, ($2528\pm20$) keV (shown by vertical dashed lines), is $r_{\rm{BG_{13}}}\approx19$ for $^{208}$Tl. The peak at $\approx2588$ keV in the $^{208}$Tl spectrum is due to escape of tellurium X-ray from TeO$_2$ detectors. (b-Inset) Reduction of background from $^{214}$Bi is $\approx53$.}
\label{fig:E-spectra-208Tl-in-screen-TeO2}
\end{figure}

\begin{figure}[!ht]
 \centering
\mbox{\epsfig{figure=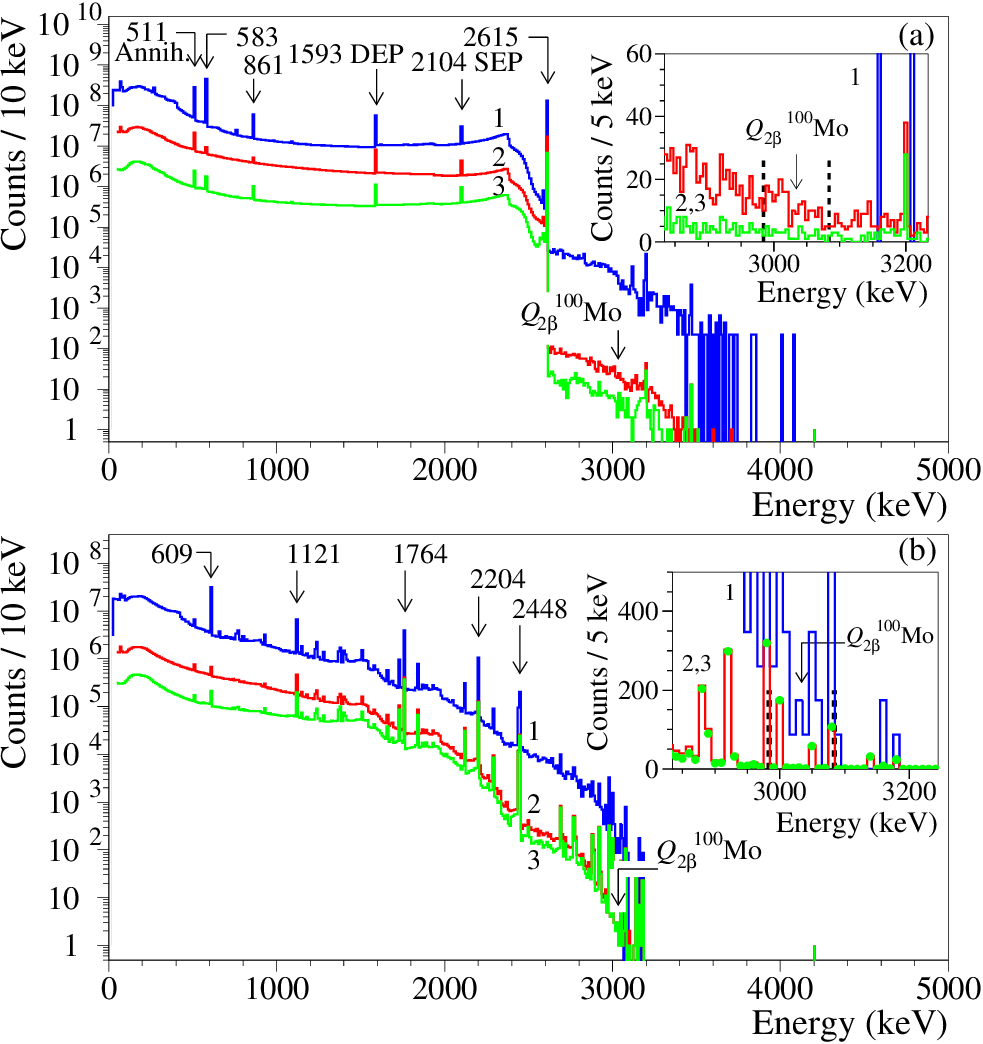,height=9.0cm}}
\caption{Monte Carlo simulated energy spectra of the Li$_2$MoO$_4$ detector array for $1.6\times10^{10}$ $^{208}$Tl (a) and $2.1\times10^9$ $^{214}$Bi (b) decays in the copper screen: (1) without scintillation veto, (2) with BGO scintillators as passive material, (3) in anti-coincidence with the scintillation veto. (Insets in panels (a) and (b)) The reduction of background with the active veto in the energy interval where a peak from $0\nu2\beta$ decay of $^{100}$Mo is expected: $r_{\rm{BG_{13}}}\approx1.1\times10^3$ for $^{208}$Tl and $r_{\rm{BG_{13}}}\approx11$ for $^{214}$Bi. For $^{214}$Bi, scintillators serve only as a passive screen, since high-energy $\gamma$ transitions mostly are not accompanied by low-energy $\gamma$ quanta.}
\label{fig:E-spectra-208Tl-in-screen-LMO}
\end{figure} 

The results of the simulations of $^{208}$Tl and $^{214}$Bi $\gamma$ quanta emitted from the copper screen and detected by the TeO$_2$ detector array are shown in Fig. \ref{fig:E-spectra-208Tl-in-screen-TeO2}. The background reduction factor $r_{\rm{BG_{ij}}}$ in the ROI is determined as:
\begin{equation}
  r_{\rm{BG_{ij}}}=\frac{BG_{i}}{BG_{j}}, 
\label{eq:BG-reduction}
\end{equation}
where $BG_i$ and $BG_j$ are the background counting rates, and indices $i$ and $j$ denote: 1 is for the detector array without the scintillation veto shield, 2 is for the scintillators as a passive material, and 3 is for the scintillation counter in anti-coincidence mode. Installation of a 5-cm layer of BGO scintillators reduces background in the energy region of $^{130}$Te $0\nu2\beta$ peak by three times, while taking into account anti-coincidences, with the scintillation counter with a threshold of 50 keV, reduces the background by a factor of $r_{\rm{BG_{13}}}\approx19$. It should be noted,  in the simulations we assume 100\% of the scintillation counter detection efficiency above the threshold.

Energy spectra simulated for the response in the Li$_2$MoO$_4$ detector array to $\gamma$ rays from $^{208}$Tl and $^{214}$Bi in the copper screen are shown in Fig. \ref{fig:E-spectra-208Tl-in-screen-LMO}. The background reduction in anti-coincidence with the scintillation veto in the energy interval $(3034 \pm 50)$ keV is $r_{\rm{BG_{13}}}=1.1(2)\times 10^3$ for $^{208}$Tl. Since $\gamma$-ray transitions in $^{214}$Bi with emission of $\gamma$ rays above 2.9 MeV mostly proceed to the ground state of the daughter without emission of other $\gamma$-rays, such high energy $\gamma$-rays are not rejected by the active veto in case of their full absorption in a Li$_2$MoO$_4$ detector. Therefore, for $^{214}$Bi, the BGO scintillation veto array acts more as a passive screen for the Li$_2$MoO$_4$ detectors with a reduction of background in the ROI for $^{100}$Mo of $r_{\rm{BG}_{13}} \approx 11$. However, it is essential that the active screen dramatically reduces the background caused by its own radioactive contamination.

\begin{figure}[!ht]
 \centering
\mbox{\epsfig{figure=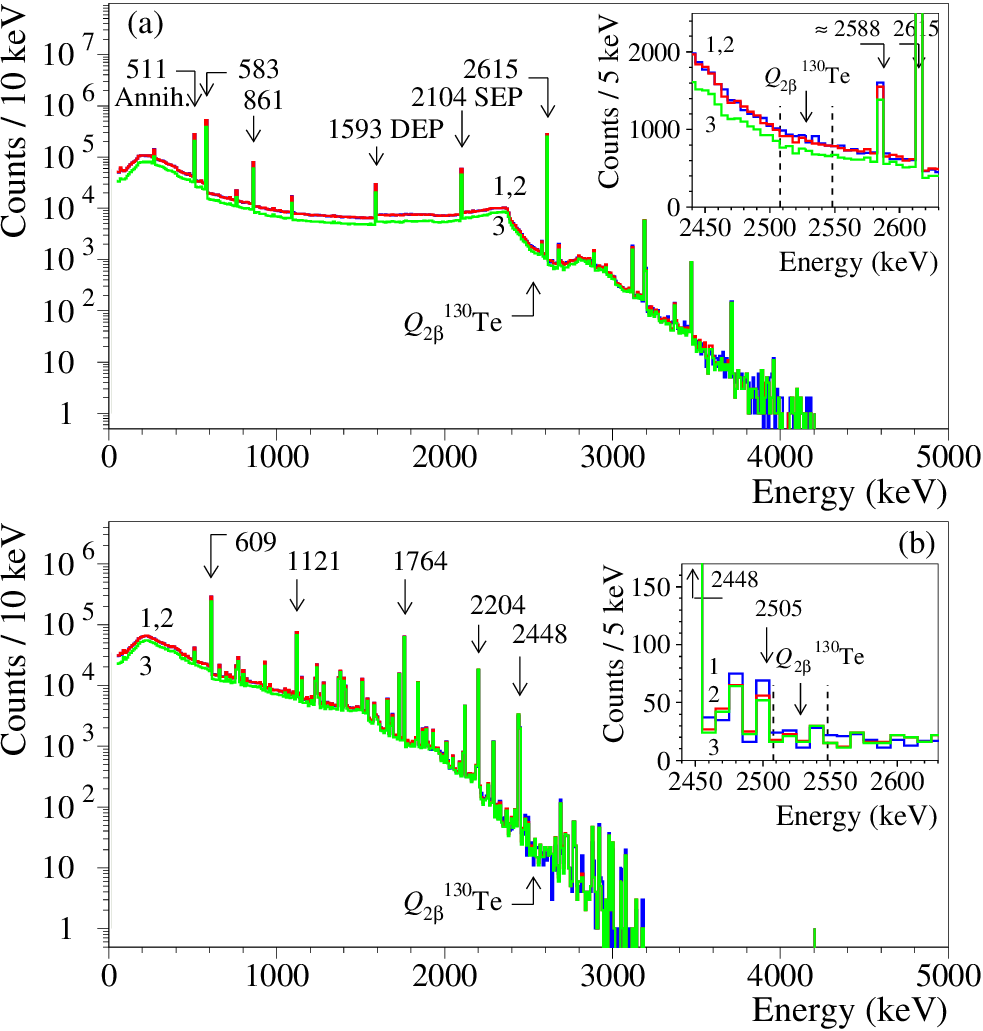,height=9.0cm}}
\caption{Monte Carlo simulated energy spectra of the TeO$_2$ detector array for $2.4\times10^6$ $^{208}$Tl (a) and $2.4\times10^6$ $^{214}$Bi (b) decays in the copper crystal holders: (1) without scintillation veto, (2) with BGO scintillators as passive material, (3) in anti-coincidence with the scintillation veto. (Insets in panels (a) and (b)) The reduction of background with the active veto in the energy interval where a peak from $0\nu2\beta$ decay of $^{130}$Te is expected, ($2528\pm20$) keV, is $r_{\rm{BG_{13}}}\approx1.2$ for both $^{208}$Tl and $^{214}$Bi.}
\label{fig:E-spectra-208Tl-in-holders-TO}
\end{figure}

The results of the background simulation from $^{208}$Tl and $^{214}$Bi decays in the copper holders of the TeO$_2$ detectors are presented in Fig. \ref{fig:E-spectra-208Tl-in-holders-TO}. The background reduction due to the active veto for the entire array of TeO$_2$ detectors is rather insignificant, $r_{\rm{BG_{13}}}\approx1.2$, but for the outer layer of detectors (in total 776 modules, which is almost half of the detectors) the background reduction is slightly bigger with a value of $r_{\rm{BG_{13}}}\approx1.5$ for $^{208}$Tl and $r_{\rm{BG_{13}}}\approx1.4$ for $^{214}$Bi. This is important because the anti-coincidence efficiency to suppress background between the crystals in a detector array is usually lowest in the outer layer. 
The simulated spectra from $^{208}$Tl and $^{214}$Bi decays in the copper holders detected by the Li$_2$MoO$_4$ detector array are similar to those in the TeO$_2$ detectors, so we do not present them here.

\begin{figure}[!ht]
 \centering
\mbox{\epsfig{figure=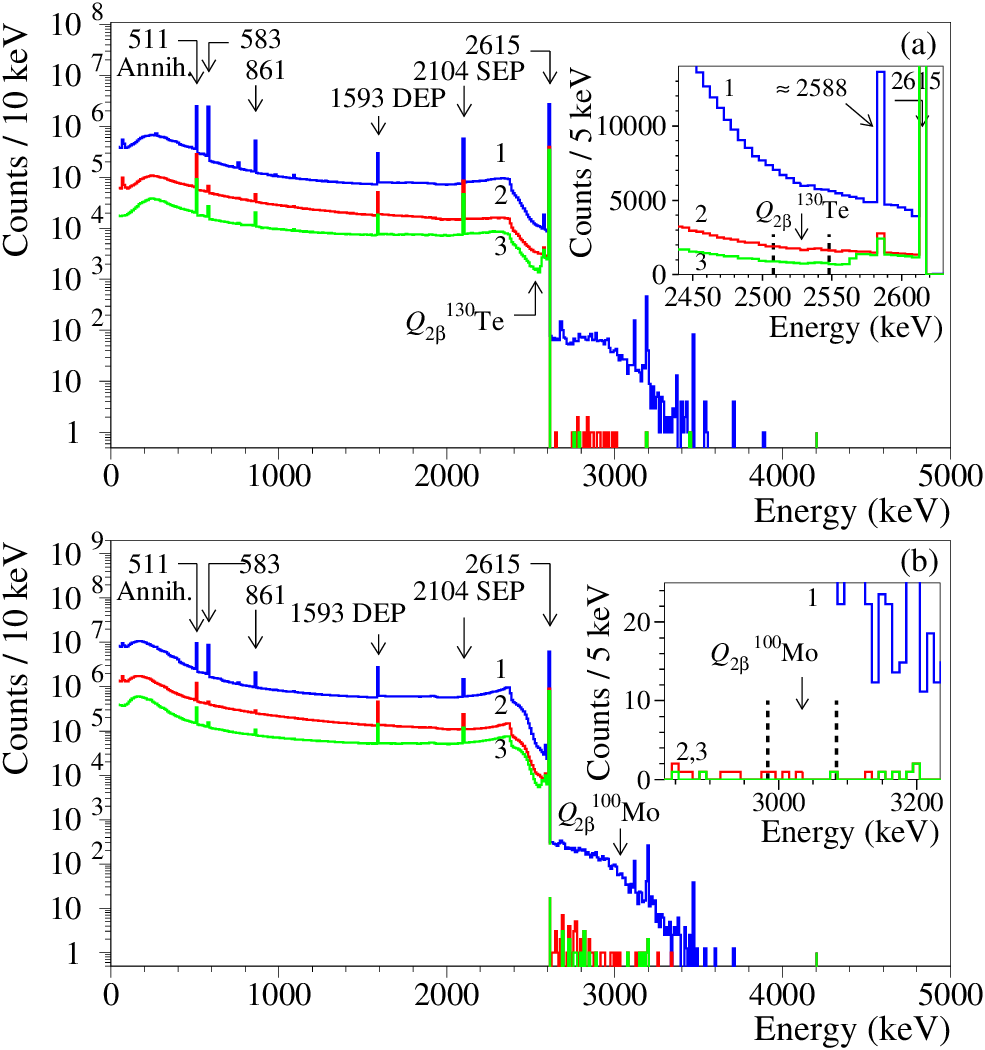,height=9.0cm}}
\caption{Monte Carlo simulated energy spectra of the TeO$_2$ (a) and the Li$_2$MoO$_4$ (b) detector arrays for $3.6\times 10^8$ and $3.0\times 10^9$ $^{208}$Tl decays, respectively, in the lead shield: (1) without scintillation veto, (2) with BGO scintillators as passive material, (3) in anti-coincidence with the scintillation veto. (Insets in panels (a) and (b)) The reduction of background with the active veto in the ROI for $^{130}$Te and $^{100}$Mo is $r_{\rm{BG_{13}}}\approx7$ for TeO$_2$ and $r_{\rm{BG_{13}}}\approx190$ for Li$_2$MoO$_4$. }
\label{fig:E-spectra-208Tl-in-Pb-TO}
\end{figure}

The results of simulations of the response of TeO$_2$ and Li$_2$MoO$_4$ detector arrays to $^{208}$Tl in the lead shield are shown in Fig. \ref{fig:E-spectra-208Tl-in-Pb-TO}. In TeO$_2$ detectors, the background reduction due to the active veto is $r_{\rm{BG_{13}}}\approx7$, while in Li$_2$MoO$_4$ detectors, the background reduction factor is $r_{\rm{BG_{13}}} \approx 190$. $^{214}$Bi was not simulated in the lead because it requires a very long time to calculate.

A summary of the background reduction factors for TeO$_2$ and Li$_2$MoO$_4$ detector arrays is given in Table \ref{tab:BG-reduction}.

\begin{table*}
\caption{Monte-Carlo simulated background reduction factors ($r_{\rm{BG_{12}}}$ and $r_{\rm{BG_{13}}}$) in the energy interval around the decay energy $Q_{2\beta}$ of $^{130}$Te ($2528\pm20$) keV for the TeO$_2$ detector array and in the energy interval around the decay energy $Q_{2\beta}$ of $^{100}$Mo ($3034\pm50$) keV for the Li$_2$MoO$_4$ detector array surrounded by a BGO scintillation veto with the energy threshold 50 keV, assuming contamination of the setup materials by $^{208}$Tl and $^{214}$Bi. The standard statistical uncertainties of $r_{\rm{BG_{12}}}$ and $r_{\rm{BG_{13}}}$ are given in parentheses.}
\begin{tabular}{llllll}
\hline\noalign{\smallskip}
Material   	& Nuclide 	& \multicolumn{2}{c}{TeO$_{2}$} & \multicolumn{2}{c}{Li$_2$MoO$_4$} \\
\cline{3-6}
 ~          & ~        & $r_{\rm{BG_{12}}}$   & $r_{\rm{BG_{13}}}$ &$r_{\rm{BG_{12}}}$   & $r_{\rm{BG_{13}}}$ \\
\noalign{\smallskip}\hline\noalign{\smallskip}
Cu screen 	& $^{208}$Tl & 2.81(2) & 18.9(4) & 250(17) & $1.1(2)\times10^3$ \\
 ~		   & $^{214}$Bi & 31(3) & 53(6)   & 11.0(6) & 11.3(6) \\
Pb shield 	& $^{208}$Tl & 3.64(2) & 7.11(6) & 129(58) & 192(97) \\
Cu holders 	& $^{208}$Tl & 1.04(2) & 1.24(2) & 0.95(5) & 1.21(6) \\
~           & $^{214}$Bi & 1.11(16) & 1.17(17) & 1.4(6) & 1.4(6) \\
\noalign{\smallskip}\hline
\end{tabular}
\label{tab:BG-reduction}
\end{table*}

\subsection{Dependence on scintillation material and energy threshold}
\label{choice-of-scintillator}


Monte Carlo simulations to study the effectiveness of different scintillator candidates for the active veto around the TeO$_2$  detector array were performed; results corresponding to two assumed energy thresholds of thermal detectors, 200~keV and 50 keV, are presented in Table \ref{tab:different-scintillators}. 
The study shows a clear improvement of the scintillation veto efficiency with a decrease of the scintillation counter energy threshold, and by utilization of scintillation material with high density and atomic number, like ZnWO$_4$, BGO, or PbWO$_4$. The use of these scintillation crystals reduces the TeO$_2$-detector-array background by more than a factor 15, assuming an energy threshold of the scintillation veto counters of 50 keV.

\begin{table*}
\caption{Background reduction factors ($r_{\rm{BG_{13}}}$, see text) in the energy interval around the decay energy $Q_{2\beta}$ of $^{130}$Te ($2528\pm20$) keV for the TeO$_2$ detector array surrounded by scintillation veto made of different scintillation materials, with the energy threshold 200 keV and 50 keV, assuming contamination of the low-temperature copper screen by $^{208}$Tl. Data on density, effective atomic number ($Z_{\rm{eff}}$), and scintillation efficiency at low temperatures for different crystal scintillators are given for comparison. The standard statistical uncertainties of $r_{\rm{BG_{13}}}$ are given in parentheses. }
\label{tab:different-scintillators}       
\begin{tabular}{llllll}
\hline\noalign{\smallskip}
Scintillation   & Density & $Z_{\rm{eff}}$ & Scintillation & \multicolumn{2}{c}{$r_{\rm{BG_{13}}}$} \\
\cline{5-6}
material        & (g/cm$^3$) & ~ &  efficiency   &    \\
~ & ~ & ~ & (photons/keV) & 200 keV & 50 keV \\
\noalign{\smallskip}\hline\noalign{\smallskip}
NaI             & 3.67 & 51 & (25--26) at 1.7 K \cite{Sailer:2012} & 2.85(7) & 6.3(2) \\
CaMoO$_4$       & 4.20 & 20 & (15--25) at 7 K \cite{Mikhailik:2007a} & 3.02(8) & 7.7(3) \\
CsI             & 4.51 & 54 & 107 at 77 K \cite{Moszynski:2003} & 3.47(9) & 8.0(3) \\
CaWO$_4$        & 6.10 & 62 & (16--26) at 20 mK \cite{Moszynski:2005,Mikhailik:2007b} & 4.9(2) & 13.8(7) \\
LYSO:Ce         & 7.40 & 65 &  (16--18) at 4 K \cite{Pankratova:2024} & 6.0(2) & 19(1) \\
BGO             & 7.13 & 75 & 31 at 77 K \cite{Moszynski:2003} & 6.3(2) & 19(1) \\
                &    &  &  23 at 77 K \cite{Moszynski:2004}    &  &  \\
                &      &    & 24 at 6 K \cite{Gironnet:2008}  &  & \\
ZnWO$_4$        & 7.87 & 61 & (18--30) at 8 K \cite{Kraus:2009} & 6.3(2) & 20(1) \\
PbWO$_4$        & 8.28 & 73 &  (4--7) at 7 K \cite{Danevich:2010}  & 8.6(4) & 28(2) \\
\noalign{\smallskip}\hline
\end{tabular}
\label{tab:BG-reduction-factors}
\end{table*}

\begin{figure}[!ht]
 \centering
\mbox{\epsfig{figure=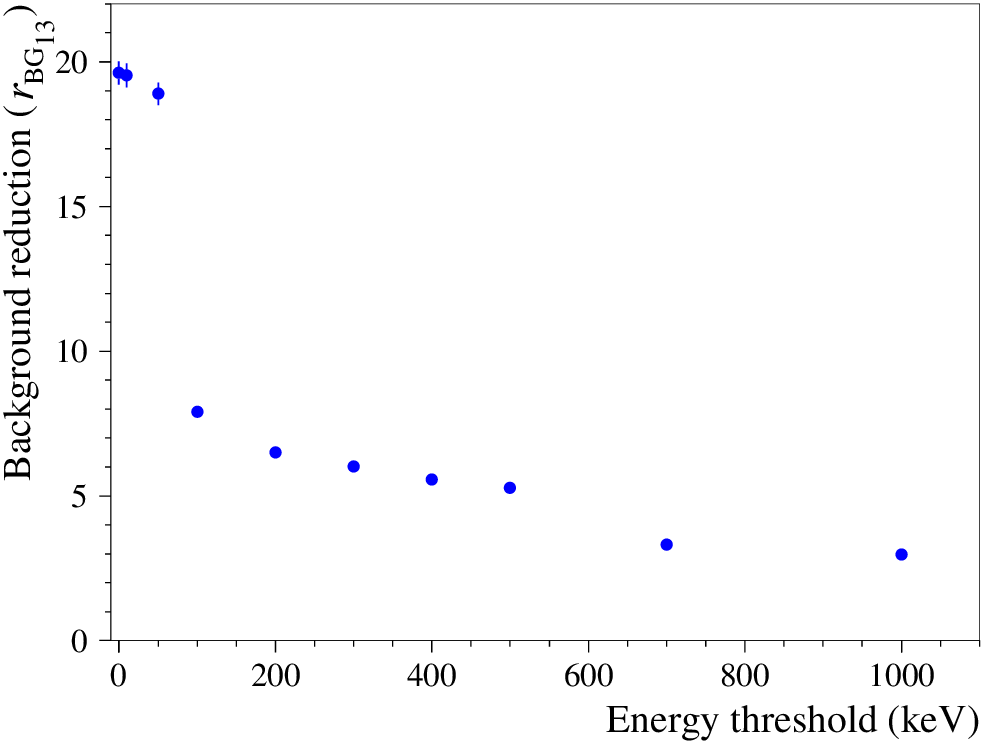,height=6.0cm}}
\caption{Dependence of BGO scintillation veto efficiency (expressed in background reduction factor $r_{\rm{BG}_{13}}$) on the energy threshold of a BGO scintillation counter for suppression of background of the TeO$_2$ detector array due to decays of $^{208}$Tl in the copper screen. It can be seen that below the threshold of $\approx100$ keV, the efficiency of the scintillation veto increases sharply.}
\label{fig:E-threshold}
\end{figure}

A more fine-grained evaluation of the impact of the energy threshold on the veto efficiency has been performed for the BGO based active veto and is shown in Fig. \ref{fig:E-threshold}. It can be seen that below the threshold of $\approx 100$ keV, the efficiency of the scintillation veto increases sharply. As Monte Carlo simulations show, it arises from the rejection of the 2615-keV $\gamma$-ray quanta, which, after losing some energy, preferentially inside scintillators, and being detected by the veto, do not fall within the ROI of $^{130}$Te.

\subsection{Simulations of photons propagation}
\label{sec:sim-photons}

One can see in Fig. \ref{fig:E-threshold} that an energy threshold of the scintillation counter of less than $\sim50$ keV is required to reduce the external background effectively. We have performed Monte Carlo simulations of scintillation photon propagation to estimate the requirements for the size and characteristics of BGO scintillation elements and photodetectors to provide high enough signal-to-noise ratio, necessary to achieve the low energy threshold of the scintillation counters. As briefly discussed below, BGO appears as a promising material for cryogenic scintillation veto applications. 

BGO scintillators were developed and studied in the mid-1970s and early 1980s. The material exhibits high detection efficiency for $\gamma$ quanta, high scintillation efficiency at low temperatures, and low radioactive contamination by uranium and thorium, along with their daughters \cite{Mcgregor:2018scintillators,Danevich:2018}. A key point is that the production of large-volume crystal scintillators (tens of centimetres in length) has been established (e.g., 40-cm-long BGO crystals are described in \cite{Gusev:2001}, while investigations of 60-cm-long crystals are reported in \cite{Ji:2014}). 

To simulate the propagation and absorption of scintillation photons in a scintillation counter, data on the scintillation (scintillation efficiency, emission spectra) and optical (index of refraction, optical transmission) properties of BGO scintillators at low temperatures are needed. Further modelling of photon absorption in photodetectors requires knowledge of the properties and characteristics of photodetectors that convert a light signal into an electrical pulse.  

The scintillation efficiency of BGO ($N_{\rm{ph}}$), ranging from 6 to 9 photons per keV at room temperature, is reported in literature \cite{Holl:2002,Humm:2003,Gironnet:2008}. For comparison, it has been measured as (23--31) photons per keV at liquid nitrogen temperature \cite{Moszynski:2003,Moszynski:2004} and $N_{\rm{ph}}=24$ photons per keV at 6 K \cite{Gironnet:2008}, allowing us to adopt the latter value for simulations of a scintillation counter at milli-Kelvin temperatures.

\begin{figure}
\centering
\mbox{\epsfig{figure=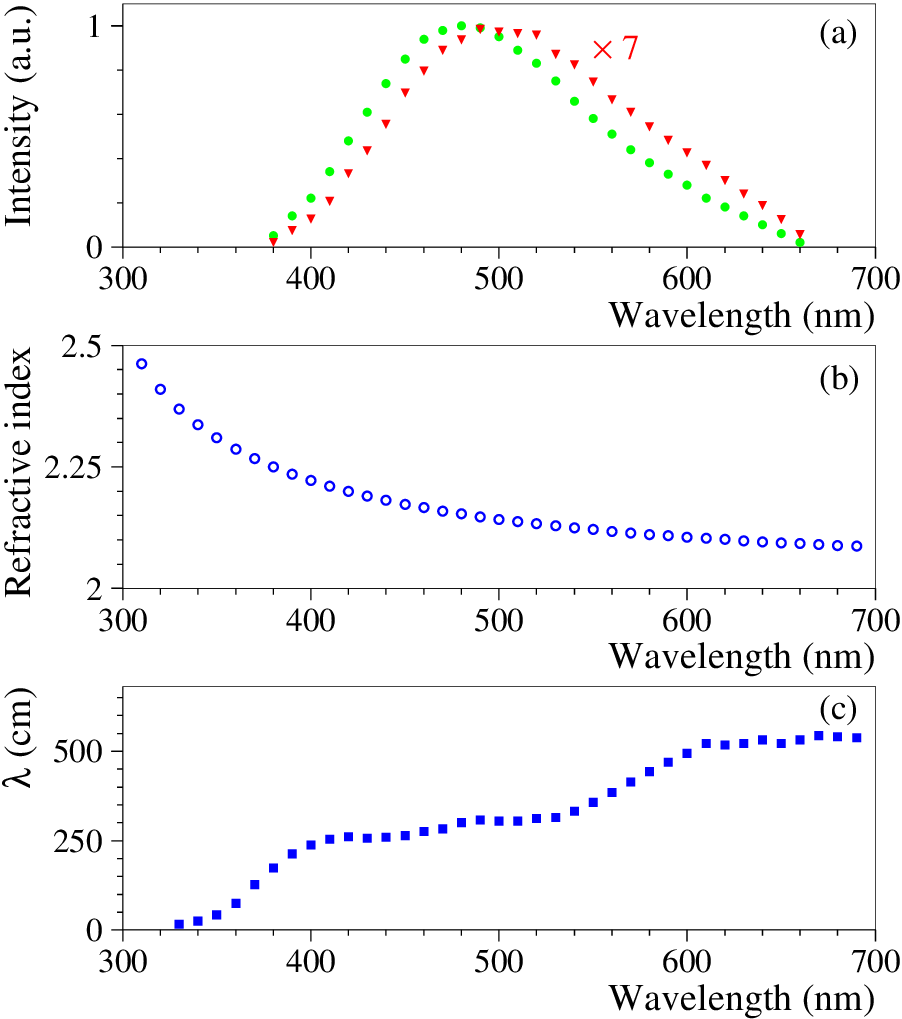,height=8.0cm}}
\caption{Scintillation emission spectra (a): primary emitted by a BGO scintillator (green dots) \cite{Melcher:1985}, and absorbed in a Ge photodetector (simulated by Monte Carlo method, red triangles; the distribution is multiplied by 7 for better visibility). An average energy of emitted photons is $E_{em}=2.51$ eV, while an average energy of absorbed photons is slightly smaller, $E_{ab}=2.43$ eV due to a worse transmittance of BGO crystals at lower wavelength. Dependence of BGO crystals' refractive index on wavelength \cite{Jellison:2010} (b), absorption length ($\lambda$) of BGO calculated using the data from \cite{Yuan:2018} (c).}
 \label{fig:emis-ref-transp}
\end{figure}

BGO scintillators emit light in the range 380--660 nm with a maximum at $\approx 480$~nm \cite{Melcher:1985} (see Fig. \ref{fig:emis-ref-transp} (a)). An average energy of emitted photons can be estimated as $E_{em}=2.51$~eV. As shown in \cite{Verdier:2011}, the scintillator emission spectrum remains practically unchanged down to 3 K. Thus, we use the emission spectrum from \cite{Melcher:1985}.  

The refractive index of BGO has been reported as 2.14 at 500 nm \cite{Weber:1973} and 2.15  at 480 nm \cite{Melcher:2000}. The dependence on wavelength was measured in \cite{Jellison:2010} and is reported in Fig. \ref{fig:emis-ref-transp} (b) as used in our simulation study.

BGO exhibits a rather high optical transparency at room temperature (see e.g., \cite{Jellison:2010,Sasano:2013,Ji:2014,Brunner:2017,Yuan:2018}). For instance, in work \cite{Sasano:2013} the absorption length of BGO crystal is assumed to be 4 m at the emission maximum. There are no data on the optical transmission of BGO crystals at milli-Kelvin temperatures. However, taking into account that BGO is non-hygroscopic, has a robust crystal structure, and possible phase transitions down to 4 K were not reported, it is expected that the optical transmissivity of the crystal is unchanged down to milli-Kelvin temperatures. The dependence of the absorption length ($\lambda$) on wavelength  presented in Fig. \ref{fig:emis-ref-transp} (c) was calculated using the experimental data from \cite{Yuan:2018}. The measurement results in \cite{Yuan:2018} provide a slightly smaller absorption length, $\lambda\sim3$ m, 
than that used in \cite{Sasano:2013} ($\lambda\sim4$ m) for the photons propagation simulations. However, even the data  \cite{Yuan:2018} appear to show the leading edge of the temporary crystal production technology. The BGO crystals used in the present study, as described in Sect. \ref{results}, are less transparent. Nevertheless, the optimistic characteristics actually achieved in practice \cite{Yuan:2018} can be assumed in our simulations of possible future experiments. 

A single simplified scintillation element of such an active veto was considered for the optical photon simulation and preliminary evaluation of its possible performance using GEANT4. Our model consists of an element 100 cm long, $5\times5$ cm$^2$ in cross-section with two Ge light detectors $5 \times 5 \times 0.03$ cm$^3$, positioned at both ends of the scintillator at a distance of 0.05 cm. As pointed out in Sect. \ref{Idea-of-scint-active-veto}, bare scintillation crystals are simulated here, taking into account the cryogenic veto's ability to reject surface $\alpha$ and $\beta$ particles of different origin. Different crystal scintillators surface conditions were simulated: fully polished, polished with partial diffusivity, and fully diffused. The emission of scintillation photons was simulated at 99 points along the scintillator, on a line passing through the centre of the crystal cross-section, with a step of 1 cm, the extreme points were at a distance of 1 cm from the ends of the crystal. The scintillation photons were assumed to be emitted isotropically. For simulations of photon absorption, the optical properties of Ge-based light detectors were taken from \cite{Aspnes:1983,Nunley:2016}.

\begin{figure}
\centering
\mbox{\epsfig{figure=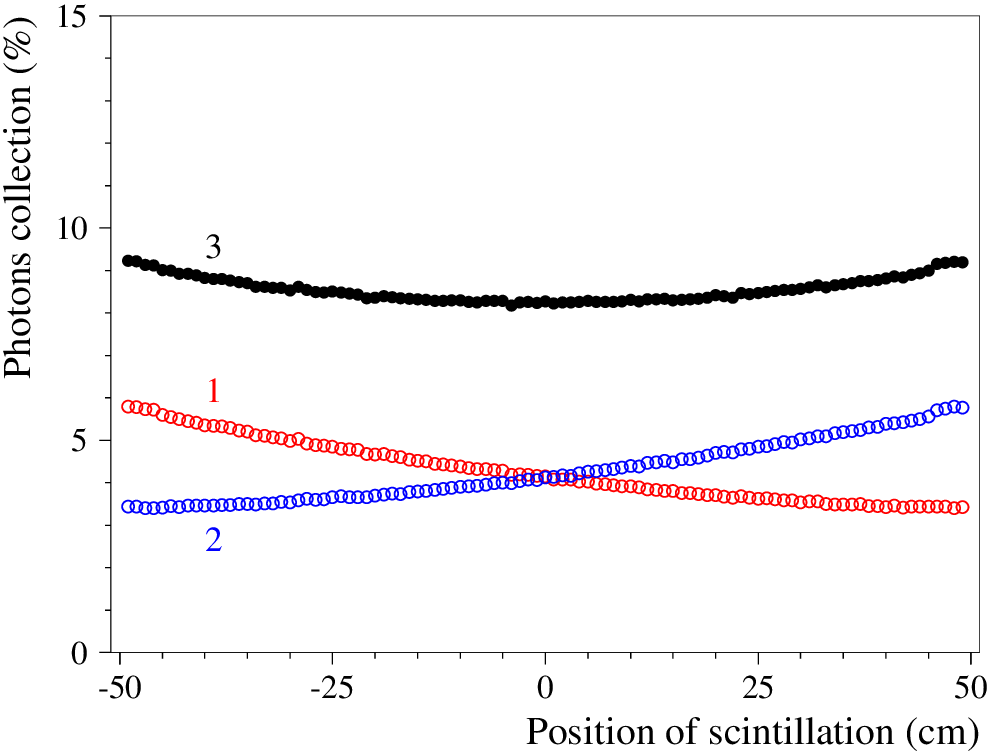,height=6.0cm}}
\caption{Dependence of scintillation photons collection on the position of scintillation point in a BGO counter (with a size of $100\times5\times5$ cm$^3$) with all the crystal surfaces polished with 10\% diffusivity, viewed by two Ge light detectors positioned at both ends of the scintillator. The dependencies for the two light detectors are denoted as ``1'' and ``2''; ``3'' is for the total collection by two photodetectors. }
 \label{fig:non-uniformity-1}
\end{figure}

\begin{figure}
\centering
\mbox{\epsfig{figure=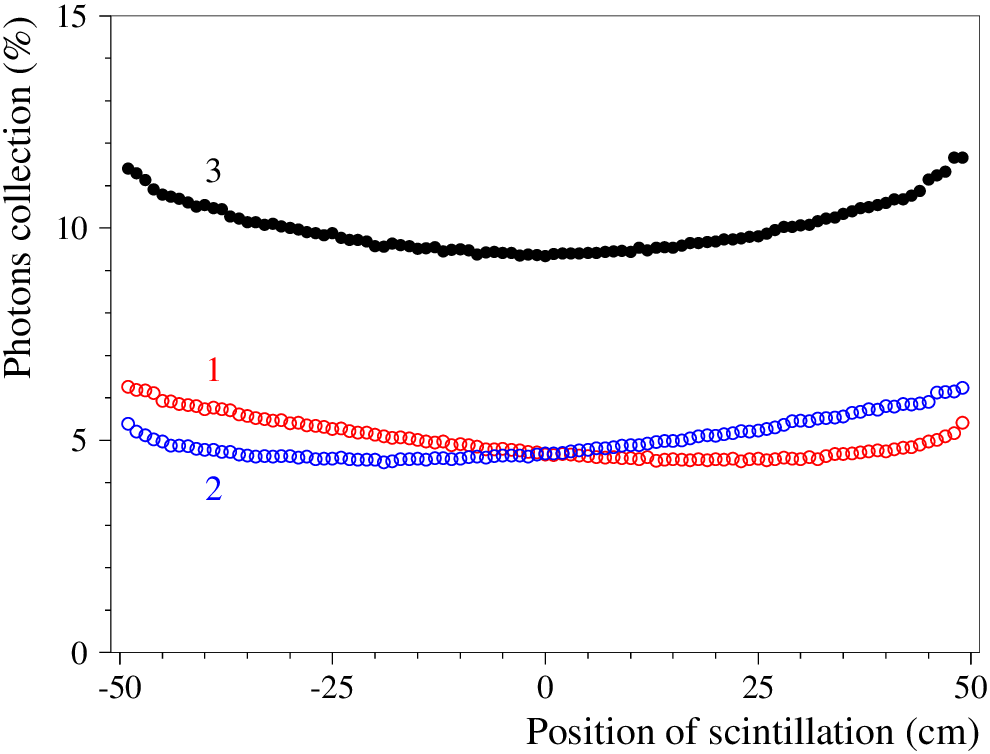,height=6.0cm}}
\caption{Dependence of scintillation photons collection on the position of scintillation point in a BGO counter ($100\times5\times5$ cm$^3$) with the side faces of the crystal polished with 10\% diffusivity, and 100\% diffusivity of the end faces, viewed by two Ge light detectors positioned at both ends of the scintillator. The dependencies for the two light detectors are labelled as ``1'' and ``2''; ``3'' is for the total collection by two photodetectors. }
 \label{fig:non-uniformity-2}
\end{figure}

The simulation results for all crystal surfaces polished with 10\% diffusivity are shown in Fig. \ref{fig:non-uniformity-1}, while the results for 10\% diffusivity of the side faces, and 100\% diffusivity of the end faces are presented in Fig. \ref{fig:non-uniformity-2}. The parameter ``Photons collection'' means the percentage of photons absorbed inside the Ge light detectors over the photons emitted in the BGO scintillator. It can be seen that the light collection from the scintillator is slightly non-uniform. 
The non-uniformity in light collection 
can be attributed to the light's path in combination with total internal reflection. It can be expressed through non-uniformity coefficient $\xi$ as a ratio of the maximal light-collection value ($LC_{\rm{max}}$) to the minimal one ($LC_{\rm{min}}$):
\begin{equation}
  \xi=\frac{LC_{\rm{max}}}{LC_{\rm{min}}}. 
\label{eq:LC-non-uniformity}
\end{equation}
The non-uniformity for the scintillation counter configuration presented in Fig. \ref{fig:non-uniformity-1} is $\xi=1.12$. A better overall light collection is achieved with the second optical configuration of the counter (see Fig. \ref{fig:non-uniformity-2}), albeit with a slightly worse $\xi=1.25$. 

\begin{figure*}
\centering
\mbox{\epsfig{figure=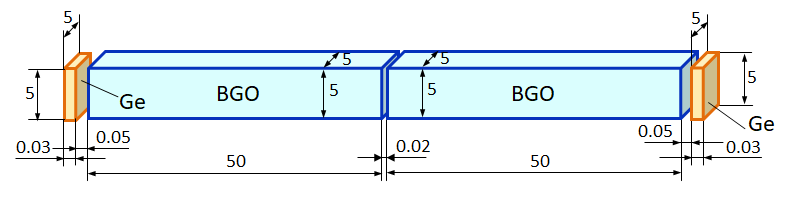,height=3.8cm}}
\caption{A simplified scheme of the counter with two BGO crystal scintillators ``cut'' into two equal 50-cm-long pieces with a 0.2-mm gap between the crystals, and two Ge-based light detectors (denoted as ``Ge'') viewing the scintillators from both ends, used for GEANT4 simulation of photon collection. All the sizes are in centimetres.} 
 \label{fig:counter-of-2BGO}
\end{figure*}

\begin{figure}
\centering
\mbox{\epsfig{figure=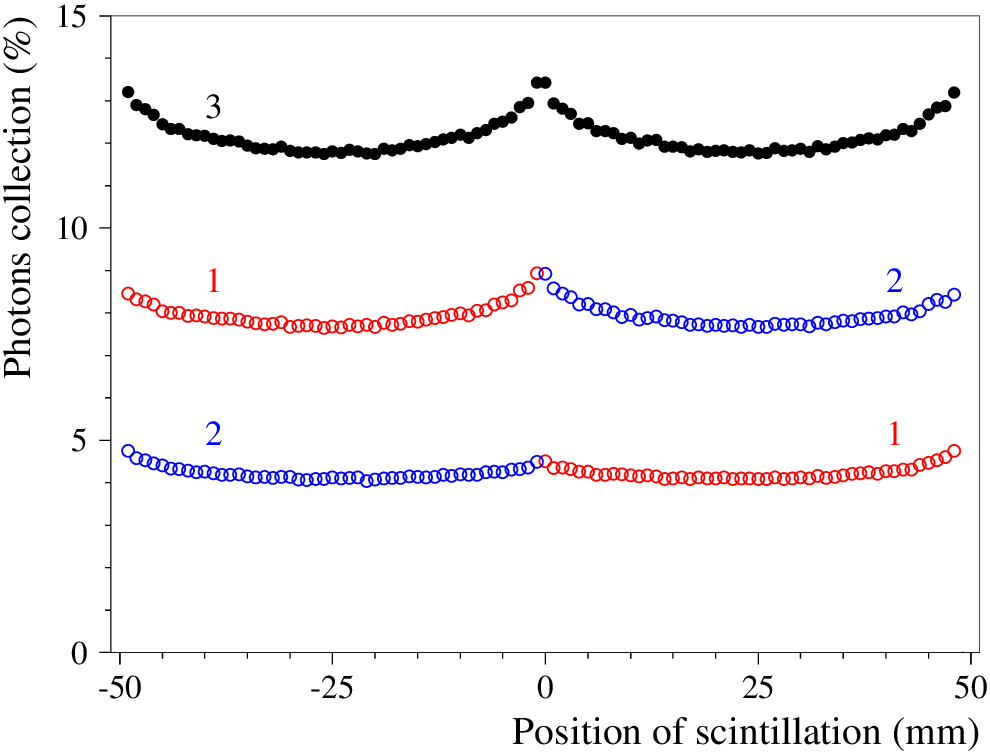,height=6.0cm}}
\caption{Dependence of scintillation photons collection on the position of scintillation point in a BGO counter ($100\times5\times5$ cm$^3$) ``cut'' into two equal 50-cm-long parts, with a 0.2-mm gap between the crystals (see Fig. \ref{fig:counter-of-2BGO}), with the side faces of the crystals polished with 10\% diffusivity, and fully diffused end faces, viewed by two Ge light detectors positioned at both ends of the scintillators. The dependencies for the two light detectors are labelled as ``1'' and ``2''; ``3'' is for the total collection by two photodetectors.} 
 \label{fig:non-uniformity-3}
\end{figure}

A substantial improvement of the light collection is obtained with crystal scintillator ``cut'' into two equal parts of 50-cm long, with a 0.2-mm gap between the crystals, with 10\% diffusivity of the side faces and fully diffused end faces. A simplified scheme of the counter is presented in Fig. \ref{fig:counter-of-2BGO}. The photons collection efficiency for the setup is shown in Fig. \ref{fig:non-uniformity-3}. The overall light collection is improved by another $\sim20\%$ at a non-uniformity comparable to the option of 100-cm long scintillators ($\zeta=1.13$). This also has considerable practical advantages. Shorter crystals are more readily available, less prone to stress related issues and more cost-effective to produce. On the other hand, the possibility of using two crystals will allow for increasing the total length of the scintillation counters, thereby increasing the volume of the $2\beta$ detector array in the setup. This is important in view of the discussed projects of multi-tonne $0\nu2\beta$ experiments. However, to assemble the two crystals, slightly more passive materials will be required, potentially increasing the radioactive background.

Using the results of the light-collection simulations, it is possible to estimate the threshold that can be achieved with BGO scintillation counters using Ge bolometric light detectors. Assuming a baseline noise level (Root Mean Square, $RMS_{\rm{bsl}}$) of 0.076 keV \cite{Alfonso:2023}, the threshold of the light detectors is typically set at a level of $\sim5~RMS_{\rm{bsl}}$, $E_{\rm{th}}^{\rm{LD}}=0.38$~keV in order to limit spurious triggers on noise excursions. An achievable energy threshold of a scintillation counter can be estimated with the following formula: 
\begin{equation}
  E_{\rm{th}}^{\rm{SC}}= \frac{ E_{\rm{th}}^{\rm{LD}} }{  N_{\rm{ph}} ~ E_{\rm{ab}} ~ LC_{\rm{min}} }. 
\label{eq:E_threshold}
\end{equation}
Considering the minimal light collection from the BGO counter $100\times5\times5$ cm$^3$ ``cut'' into two equal 50-cm-long parts with a 0.2-mm gap between the crystals is $LC_{\rm{min}}=0.12$, we obtain an energy threshold of the BGO counter instrumented with two ``ordinary'' Ge bolometric light detectors $E_{\rm{th}}^{\rm{SC}}\approx 55$ keV. 

An antireflective coating of cryogenic light detectors (that was not taken into account in the scintillation photons simulations) can be utilised to improve light collection by tens of percent \cite{Mancuso:2014}. Moreover, taking into account the already achieved characteristics of Neganov-Trofimov-Luke (NTL) light detectors with voltage-driven thermal signal amplification \cite{Novati:2019,CROSS_Run9:2026}, a several times lower energy threshold can be achieved which is expected to fully satisfy the requirements to the scintillation-veto energy threshold.

\section{Cryogenic active veto prototype}
\label{results}

Building on the veto concept and the simulations discussed in the previous section, we now turn to its experimental implementation. From a design perspective, a realistic veto (as opposed to the simplified ones discussed above) consists of long trapezoidal crystals arranged in one or two floors (depending on their length) to form the lateral part of the system, providing nearly continuous coverage around the detector array. In addition, the top and bottom parts of the veto are formed using a combination of crystals with different shapes in order to achieve 4$\pi$ coverage of the main detector array. Each veto crystal is coupled to a thin Ge wafer operated as a bolometric light detector for scintillation readout.

In this section, we present the development and characterization of a new small-scale cryogenic veto prototype based on this design and discuss their experimental performance at milli-Kelvin temperatures. The measurement provides a first successful validation of the veto concept and represents a key step toward its implementation in the Mini-BINGO demonstrator \cite{Armatol:2024}.

It is the evolution of a first prototype, described in \cite{Armatol:PhD,Khalife:2023}, which consisted out of two trapezoidal shape BGO crystals (20 cm$^2$ $\times$ 12 cm) and four Ge light detectors placed above and below each crystal to detect the scintillation light. 
However, the first prototype revealed several limitations that required improvements. The presence of large BGO crystals in the assembly significantly slowed down the cooling process of the cryostat and other thermally sensitive components, such as the light detectors. In addition, the measurements showed significant variations in light detector sensitivity and baseline energy resolution, which were primarily influenced by the strength and uniformity of the wafer’s mechanical fixation. Furthermore, light collection was not fully optimized limiting the achievable energy threshold and veto efficiency.

These observations motivated the design of a second prototype, which aimed to address these limitations by optimizing the mechanical structure, improving thermal decoupling of the light detectors, and adding reflective foils around the crystals to improve light collection.

\subsection{Prototype design}

\begin{figure}
    \centering
    \includegraphics[width=0.8\linewidth]{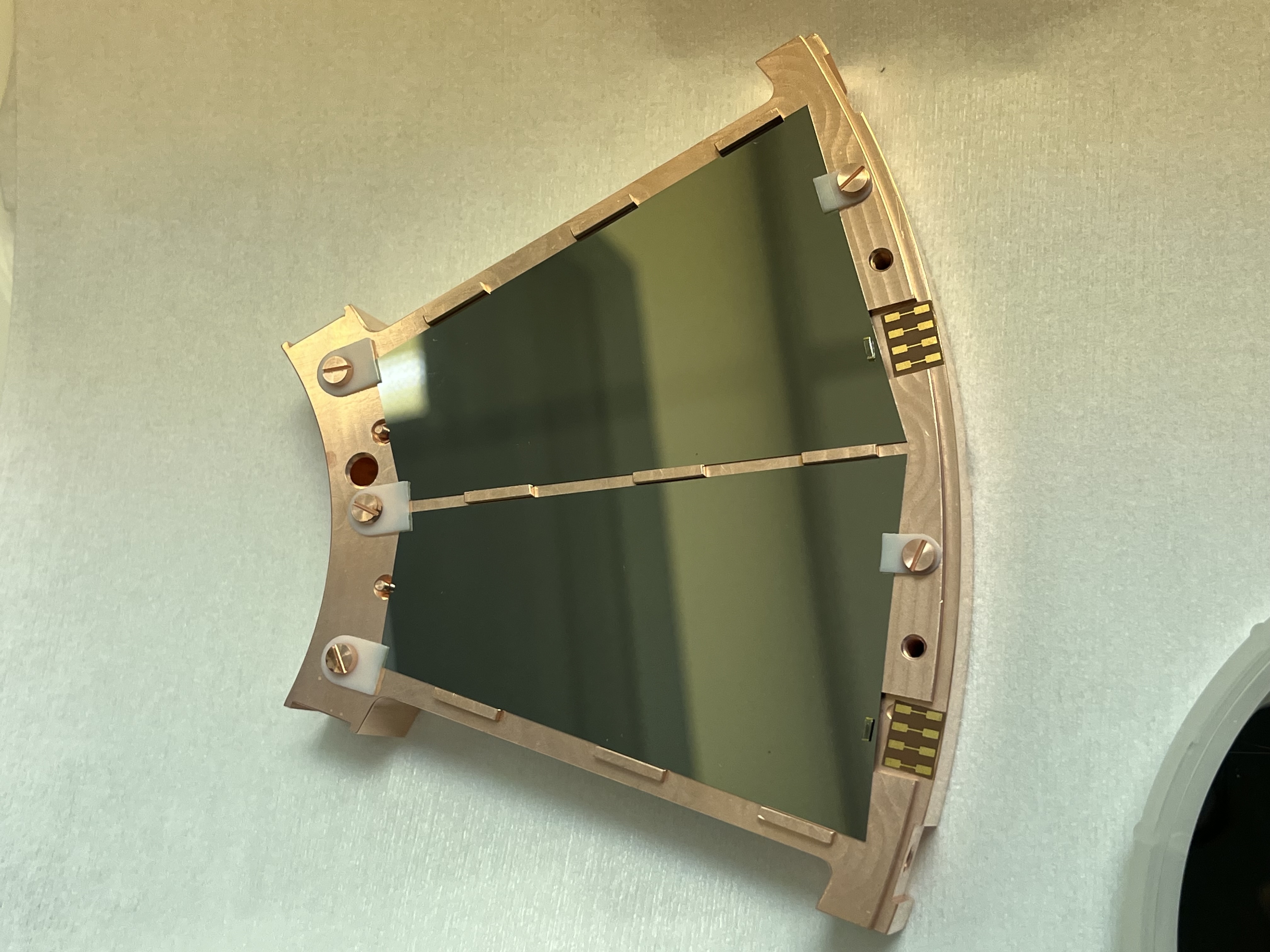}
    \caption{Trapezoidal shape germanium light detectors assembled in the copper holder and fixed with PTFE pieces.}
    \label{fig:veto2-lds}
\end{figure}

The overall concept of the second veto prototype remained similar to that of the first test. Four trapezoidal BGO crystals (BGO-1, BGO-2, BGO-3, and BGO-4), each with dimensions of $20~\text{cm}^2 \times 5~\text{cm}$, were used instead of the two 12-cm-long BGO crystals of the first prototype. This configuration provided greater flexibility to test multiple combinations of reflective materials within the limited experimental space available in an above ground cryostat at \mbox{IJCLab}, France \cite{Mancuso:2014a}. Each crystal was optically coupled to a dedicated trapezoidal Ge light detector (LD-1, LD-2, LD-3, and LD-4), labeled according to the corresponding crystal it faced (see Fig. \ref{fig:veto2-lds}). The light detectors were secured in place using three polytetrafluoroethylene (PTFE) clamps (Fig.~\ref{fig:veto2-lds}). In this configuration, the BGO crystals rested directly on the copper structure but were thermally decoupled from the light detectors, providing improved thermal isolation compared to the earlier design.

\begin{figure}
    \centering
    \includegraphics[width=0.47\linewidth]{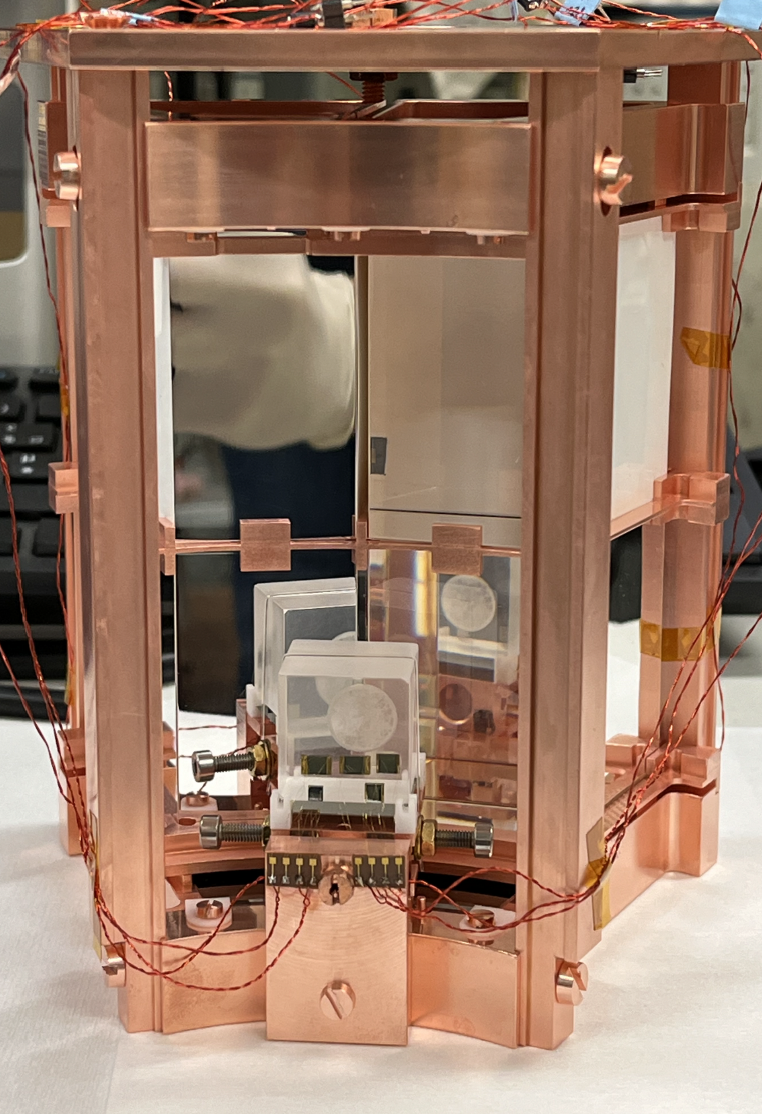}
    \includegraphics[width=0.51\linewidth]{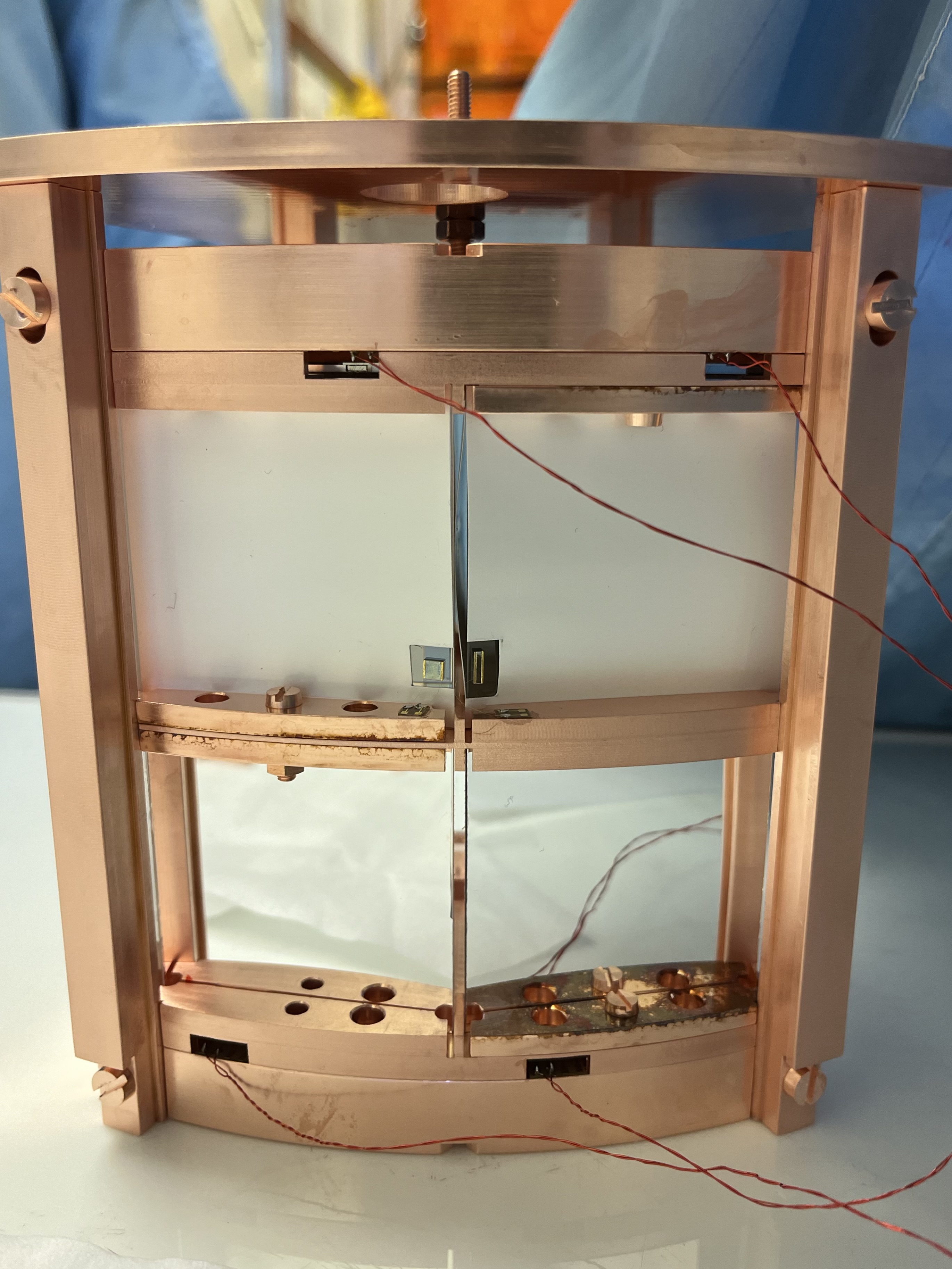}
    \caption{Photographs of the assembled second cryogenic veto prototype from the front (left) and back (right). In the left image, the BGO crystals are labelled as follows: BGO-1 (bottom left), BGO-2 (top left), BGO-3 (bottom right), and BGO-4 (top right).}
    \label{fig:tower}
\end{figure}

Similarly to the first prototype, a small BINGO module containing a $2\times2\times2$ cm$^3$ Li$_2$MoO$_4$ crystal and a  $2\times2\times0.03$ cm$^3$ high purity Ge light detector was placed facing the veto module to study coincidence events. To imitate surface contamination, a drop of liquid containing $^{234}$U and $^{238}$U isotopes was deposited on the surface of the crystal facing the BGO detectors. Since Li$_2$MoO$_4$ material is hygroscopic and the source was water-based, the drop formed a visible “crater” with the source penetrating deep into the surface of the LMO (see Fig. \ref{fig:tower} left).

Two types of reflective materials were used: Enhanced Specular Reflector (ESR) Mylar film with a thickness of 65~$\mu$m, and Melinex® 339 \cite{tekraMelinexxAEFilm} an opaque white polyester film acting as a light diffuser. The reflector material covers all sides of the BGO crystal except for the face directly facing the Li$_2$MoO$_4$ crystal. In addition, two out of the four crystals were coated on the inner side of the veto with a 200-nm-thick Al layer, providing optical isolation while being thin enough that $\alpha$ particles emitted from detector surfaces can still create a scintillation signal in the veto. Each BGO crystal in the prototype was paired with a distinct combination of reflector and coating, as summarized in Table \ref{tab:reflect}.

\begin{table}[h!]
    \caption{Summary of reflective materials and coatings used for each BGO crystal in the second veto prototype.}
    \centering
    \begin{tabular}{l|c|c}
        \hline
        Detector & Reflecting material & Aluminium coating \\
        \hline
        BGO-1 &ESR film&Yes\\
        BGO-2 &White polyester film&Yes\\
        BGO-3 &ESR film&No\\
        BGO-4 &White polyester film&No\\
        \hline
    \end{tabular}
    \label{tab:reflect}
\end{table}

Each germanium light detector was equipped with an NTD (Neutron Transmutation Doped) Ge thermistor \cite{Haller:1994} of dimensions $1 \times 1 \times 3~\text{mm}^3$. In order to monitor the cooling rate of the BGO scintillators, two of the BGO crystals (BGO-2 and BGO-4) were also instrumented with NTDs directly glued onto their surface, as shown in Fig. \ref{fig:tower} (right). Furthermore, a batch of NTDs was glued onto the small $\text{Li}_2\text{MoO}_4$ crystals for characterization purposes; however, only one of these channels was biased and used for the measurements related to the cryogenic veto performance. All NTD thermistors were attached using two-component epoxy glue Araldite\textregistered{} Rapid. In addition, a Si heater \cite{heater} was attached to the Li$_2$MoO$_4$ crystal to inject controlled energy pulses for monitoring and thermal gain drifts correction \cite{ANDREOTTI2012161}. For the electrical contacts, NTDs and heaters were bonded using gold wires to pads on a Kapton\textregistered{} foil glued onto the copper support structure. 

A photograph of the fully assembled second veto prototype, showing both the front and back sides of the module, is presented in Fig. \ref{fig:tower}.

\subsection{Detectors bolometric performance}

\begin{figure}[h!]
    \centering
    \includegraphics[width=0.98\linewidth]{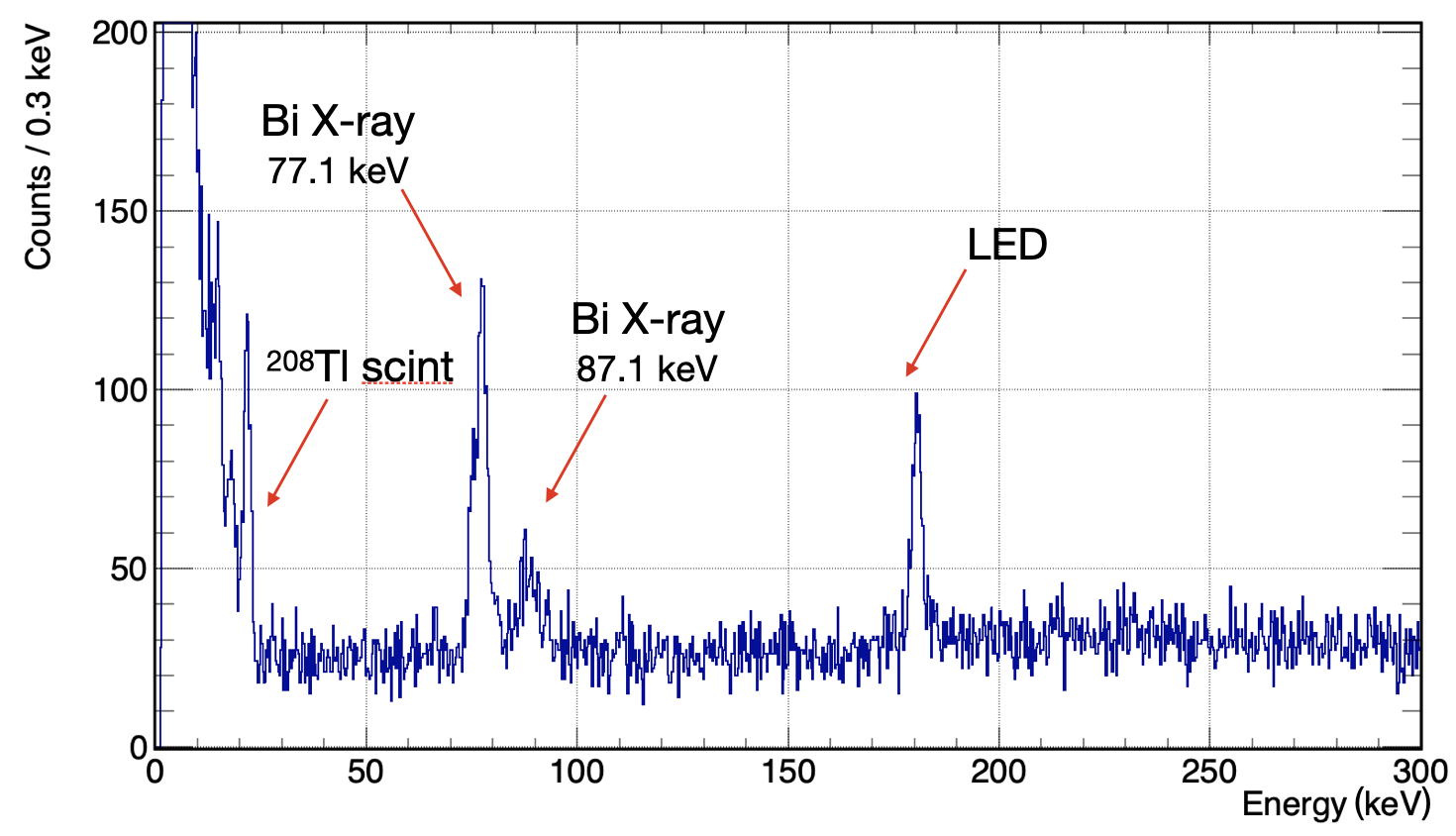}
    \caption{Energy spectrum of LD-3 acquired during 110 hours of $^{232}$Th calibration, showing Bi K$_{\alpha}$ and K$_{\beta}$ X-rays at 77.1 and 87.1 keV, respectively.}
    \label{fig:trap-ld-spectrum}
\end{figure}

A calibration of the light detectors was performed using the Bi X-ray doublet of K$_{\alpha}$ and K$_{\beta}$ at 77.1 and 87.1 keV, originating from the BGO crystals. Additionally, a cross-check of the calibration was conducted using the Mo K$_{\alpha}$ X-ray peak at 17.4~keV, which was visible in the spectra of the two bottom light detectors (LD-1 and LD-3), likely due to their proximity to the Li$_2$MoO$_4$ crystal. A typical energy spectrum is shown in Fig. \ref{fig:trap-ld-spectrum}.
The operating points and performance parameters obtained for all four trapezoidal light detectors during the 110-hour $^{232}$Th calibration run are summarized in Table \ref{tab:trap-ld-perf}.
\begin{table}[h!]
\caption{Performance of trapezoidal Ge light detectors obtained during 110 hours of $^{232}$Th calibration.}
    \centering
    \begin{tabular}{l|c|c|c|c}
        \hline
        Detector & Current & Resistance  & Sensitivity  & 
        $RMS_{\mathrm{bsl}}$\\
        ~ & (nA) & (M$\Omega$) & ($\mu$V/keV) & (keV) \\
        \hline
        LD-1 & 1.0 & 5.9 & 1.8 & 0.15\\
        LD-2 & 0.5 & 13.6 & 2.4 & 0.48\\
        LD-3 & 1.0 & 5.1 & 2.0 & 0.13\\
        LD-4 & 0.5 & 13.8 & 2.4 & 0.22\\
        \hline
    \end{tabular}
    \label{tab:trap-ld-perf}
\end{table}
The results show that LD-2 and LD-4 exhibit sensitivities comparable to LD-1 and LD-3. Although LD-2 and LD-4 operate at higher resistances, their sensitivities remain similar because the corresponding lower currents result in a comparable voltage across the thermistors. However, their baseline resolutions are significantly worse, particularly for LD-2. This degraded resolution is attributed to a higher susceptibility to noise at these operating points, rather than a deficit in intrinsic sensitivity. This behavior was confirmed during an additional $^{232}$Th calibration measurement under optimized noise conditions that slightly improved their baseline resolution, but the overall trend remained unchanged.


A dedicated noise analysis reveals that LD-1 and LD-3, both located on the bottom of the assembly, have similar and relatively low noise levels. Even though their working points were not optimized for minimum noise, their plateau levels in the 10–100 Hz region remain below 20 nV/$\sqrt{\text{Hz}}$. Conversely, LD-2 and LD-4 on the top of the module demonstrate significantly higher low-frequency noise. This is likely induced by mechanical vibrations, potentially due to suboptimal mechanical fixation by PTFE supports allowing micro-movements and increasing susceptibility to external noise.

\subsection{Surface alpha events tagging}

To assess the efficiency of the BGO-based cryogenic veto system, a coincidence analysis was performed between the signals recorded by the Li$_2$MoO$_4$ thermal detector and the surrounding trapezoidal light detectors. It effectively aims to identify overlapping signals: a heat signal in the Li$_2$MoO$_4$ crystal and a scintillation signal in the BGO, detected via its associated light detector, within the same time window (10 ms). The primary focus was placed on $\alpha$ particles emitted by the uranium source, which irradiated both the LMO and adjacent BGO crystals. For this analysis, we concentrated on BGO-1 and BGO-3, located at the bottom of the tower, which have higher geometrical coincidence efficiency and hence higher statistics to compare the impact of the Al coating on the tagging of surface events. It is important to recall that, due to the method of source deposition, the number of true surface $\alpha$ events depositing their full energy in a single BGO crystal is relatively low. More commonly, $\alpha$ particles deposit energy partially in both the Li$_2$MoO$_4$ crystal and one of the nearby BGO crystals. 

\begin{figure}[h!]
    \centering
    \includegraphics[width=0.98\linewidth]{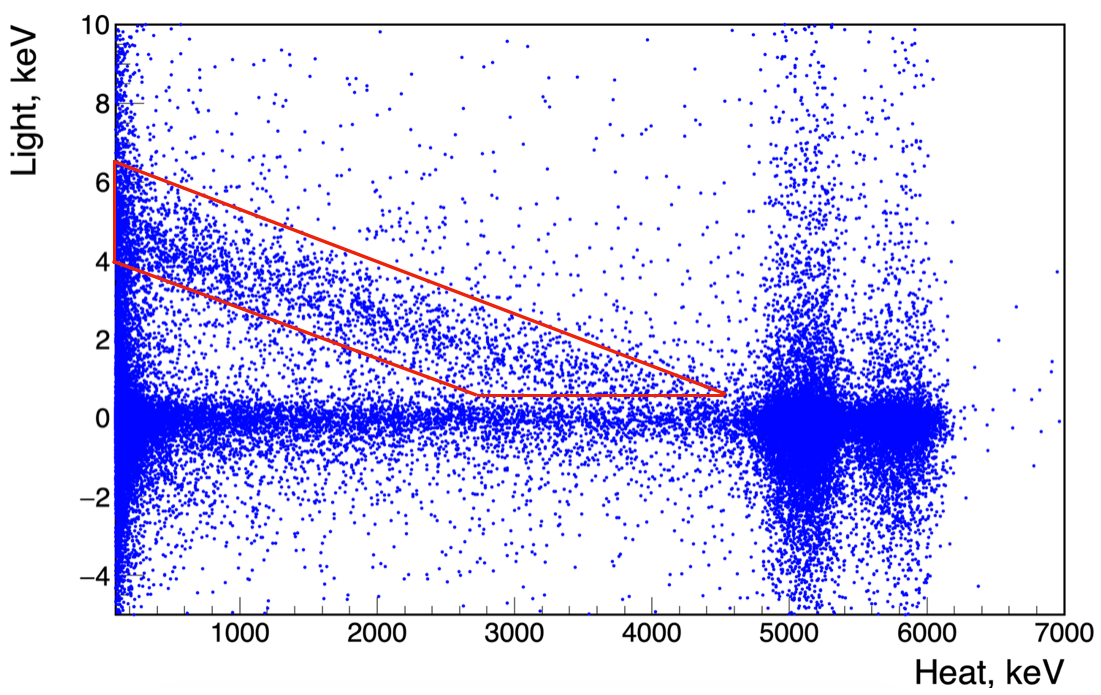}
    \includegraphics[width=0.98\linewidth]{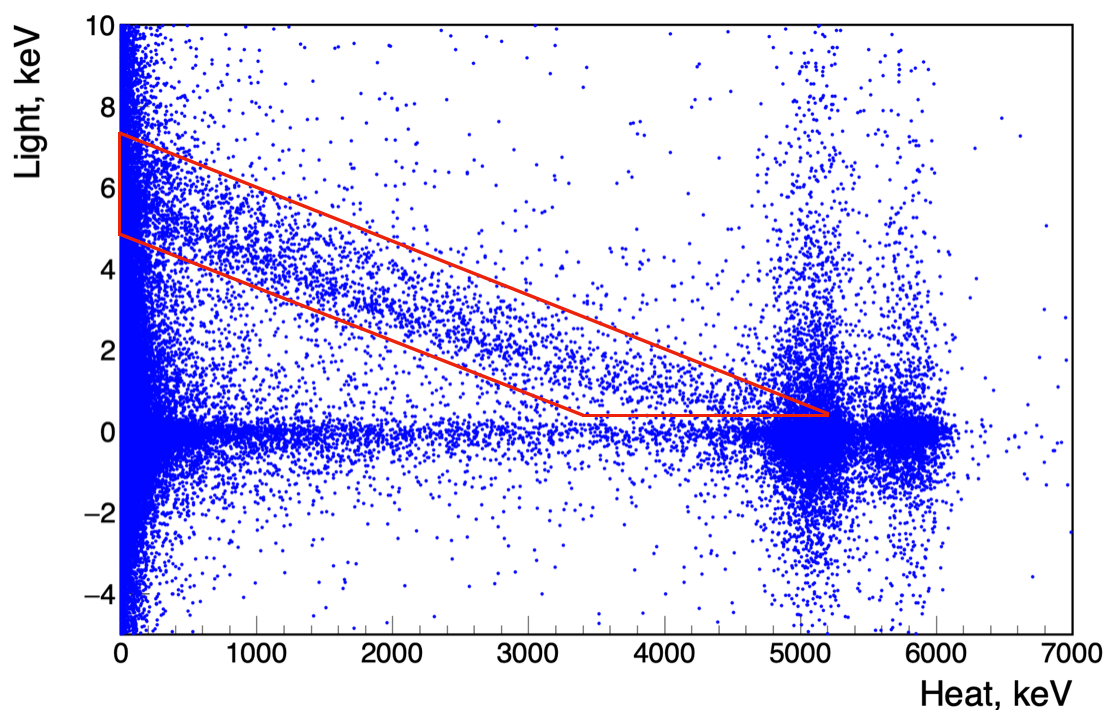}
    \caption{Scatter plots of light versus heat energy for events recorded in coincidence between the Li$_2$MoO$_4$ crystal and LD-1 (top) and LD-3 (bottom). The red frames indicate events depositing energy in both the Li$_2$MoO$_4$ and the BGO scintillators.}
    \label{fig:coincidences}
\end{figure}

The resulting scatter plots of coincident events between the Li$_2$MoO$_4$ crystal and the LD-1 and LD-3 trapezoidal light detectors are shown in Fig. \ref{fig:coincidences}. 
Two dense clusters of events can be clearly observed in the 5--6 MeV heat energy region, corresponding to the $\alpha$ peaks from the uranium source. 
Higher than nominal energies of $^{238}$U and $^{234}$U $\alpha$ peaks are due to the difference in thermal response between $\gamma$($\beta$) and $\alpha$ particles interacting inside Li$_2$MoO$_4$ cryogenic detectors (the so-called thermal quenching) \cite{Armengaud:2017}. 
These are primarily bulk $\alpha$ events in which the full energy is deposited in the Li$_2$MoO$_4$ crystal, with no coincident energy observed in the BGO scintillator. In contrast, the set of events enclosed in the red frame represents true coincidences, where energy is shared between the Li$_2$MoO$_4$ and the adjacent BGO crystal. These events are of particular interest, as they are indicative of surface or near-surface $\alpha$ interactions involving both detectors. The distribution along two diagonal bands corresponds to the two dominant $\alpha$ lines, with the energy split variably between the Li$_2$MoO$_4$ and the BGO crystals. Thanks to the relatively good energy resolution of the light detector, this group of coincident events is clearly distinguishable demonstrating the sensitivity of the BGO veto with Al coating to still tag surface backgrounds, even from short-ranged alpha particles. Furthermore, the implementation of NTL technology improves the signal-to-noise ratio of the light detector \cite{Novati:2019,CROSS_Run9:2026}, thereby enhancing both particle discrimination capability and the overall efficiency of the cryogenic active veto system.

\subsection{Light-to-heat ratio estimation}

\begin{figure}[h!]
    \centering
    \includegraphics[width=0.98\linewidth]{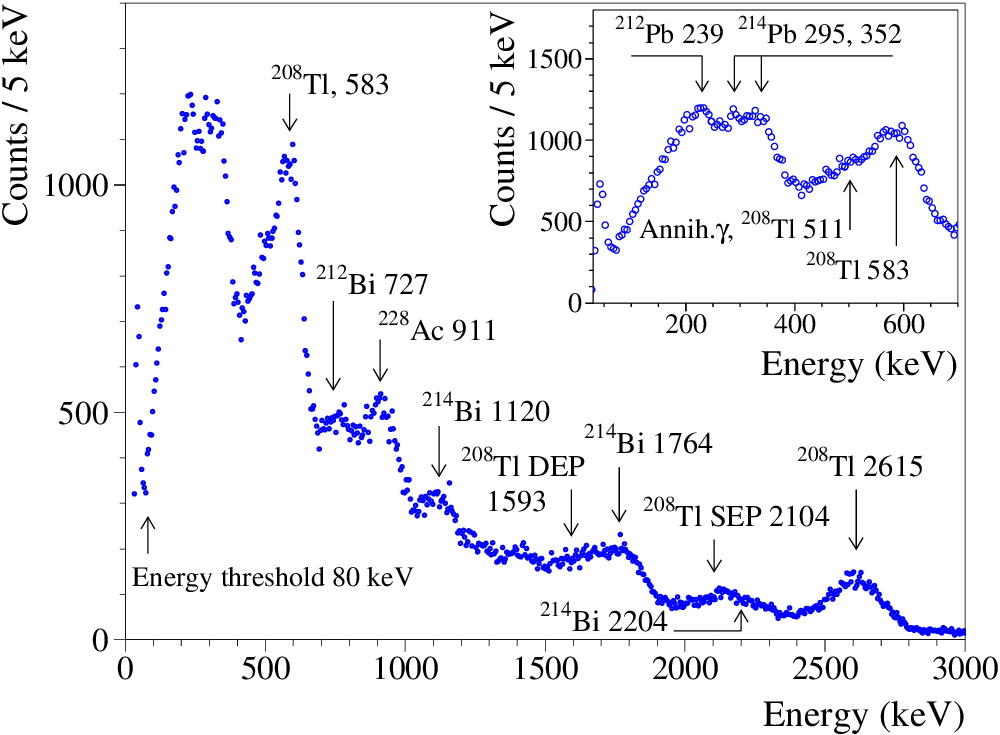}
    \caption{Energy spectrum of the LD-3 BGO counter acquired during 110 hours of calibration with a $^{232}$Th $\gamma$-ray source. The $\gamma$-ray peaks of $^{214}$Pb and $^{214}$Bi are due to the presence of radon in the setup. Energy of the $\gamma$ peaks is in keV. The hardware threshold of the scintillator is 80 keV, corresponding to $5~RMS_{\rm{bsl}}$. Genuine pulses can be triggered above this value, as shown in the energy spectrum. This threshold is consistent with optical simulations of the setup. However, in the 80--200 keV range, pulse-shape cuts are required to reject spurious pulses and noise bursts, reducing the acceptance of physical events. A safe analysis threshold, with no distortion of the energy spectrum, can therefore be set at about 200 keV. (Inset) Low energy part of the spectrum.} 
    \label{fig:bgo-ly}
\end{figure}

To estimate the light-to-heat ratio of the BGO crystals, we analyzed the low-energy region of the light detector spectra acquired during a $^{232}$Th calibration. Due to the high scintillation efficiency of BGO, crystals emit a significant amount of light when irradiated by $\gamma$ rays, and this light is detected by the corresponding Ge light detectors. As a result, the light detector spectra show a clear projection of the $^{232}$Th $\gamma$ spectrum, where the most intense $\gamma$ peaks can be readily identified, as illustrated in Fig.~\ref{fig:bgo-ly}.

Since the light detectors were previously calibrated using X-ray peaks, we can associate the measured light signal with its corresponding $\gamma$ line in the $^{232}$Th spectrum. The ratio between the light signal amplitude (in keV, measured by the light detector) and the known $\gamma$ energy deposited in the BGO crystal (in MeV) gives the light-to-heat ratio of the BGO in units of keV/MeV. This procedure was performed for each BGO-light detector pair. The obtained results are presented in Table~\ref{tab:ly-bgo}.

\begin{table}[h!]
    \caption{Measured light-to-heat ratio of BGO crystals at cryogenic temperature.}
    \centering
    \begin{tabular}{l|c|c|c}
        \hline
        Detector & Reflecting& \multicolumn{2}{c}{Light-to-heat ratio} \\
        \cline{3-4}
        ~ & material & (keV/MeV) & (photons/keV)\\
        \hline
        BGO-1 & ESR + Al& 8.1(3) & 3.3 \\
        BGO-2 & White polyester + Al& 7.4(2) & 3.0 \\
        BGO-3 & ESR& 8.4(4) & 3.5 \\
        BGO-4 & White polyester & 9.5(3) & 3.9 \\
        \hline
    \end{tabular}
    \label{tab:ly-bgo}
\end{table}

The measured values are consistent with results from prior cryogenic measurements \cite{Poda:2021hsv}, with an average light-to-heat ratio of 8.3 keV/MeV. When comparing BGO-1 and BGO-3, no significant reduction in light collection efficiency was observed due to the presence of the aluminium coating. The highest light-to-heat ratio was measured for the crystal equipped with a white polyester reflector (BGO-4), while both ESR-covered crystals showed slightly lower values. These results suggest that, at cryogenic temperatures, the choice of reflector material has only a modest impact on the overall light-to-heat ratio. Although this prevents us from identifying a clearly optimal reflector configuration, we can reasonably conclude that the aluminum coating is less detrimental than anticipated \cite{CUPID:2023dqd}. Furthermore, the coincidence analysis demonstrated that the aluminum coating does not absorb $\alpha$ particles, making it a currently acceptable option for covering the inner surfaces of the cryogenic veto.

\subsection{Simulation of light collection in the BGO-veto prototype}

Monte Carlo simulations of the photon propagation in the BGO-veto module were performed with the GEANT4 simulation package. The mechanical design of the detector was translated into an optical model of the module.

It should be noted that the optical properties and the geometry of some elements of the prototype module are not fully defined. The optical transmission data for the BGO crystal scintillator were estimated from the measurements of a BGO scintillator with a trapezoidal cross-section of $4.94 \times 4.92 \times 2.92$ cm, 23 cm long, viewed by a photomultiplier. The crystal was manufactured by the same producer using the same technology, allowing us to assume similar optical properties for the material as those used in the prototype. The scintillator was irradiated by a collimated $^{60}$Co $\gamma$ source of  at nine points evenly along the crystal. The Monte Carlo simulations using GEANT4 returned an absorption length of $\approx90$ cm at the emission maximum. The value was taken for further simulations of photon propagation in the BGO prototype.

\begin{figure}
\centering
\mbox{\epsfig{figure=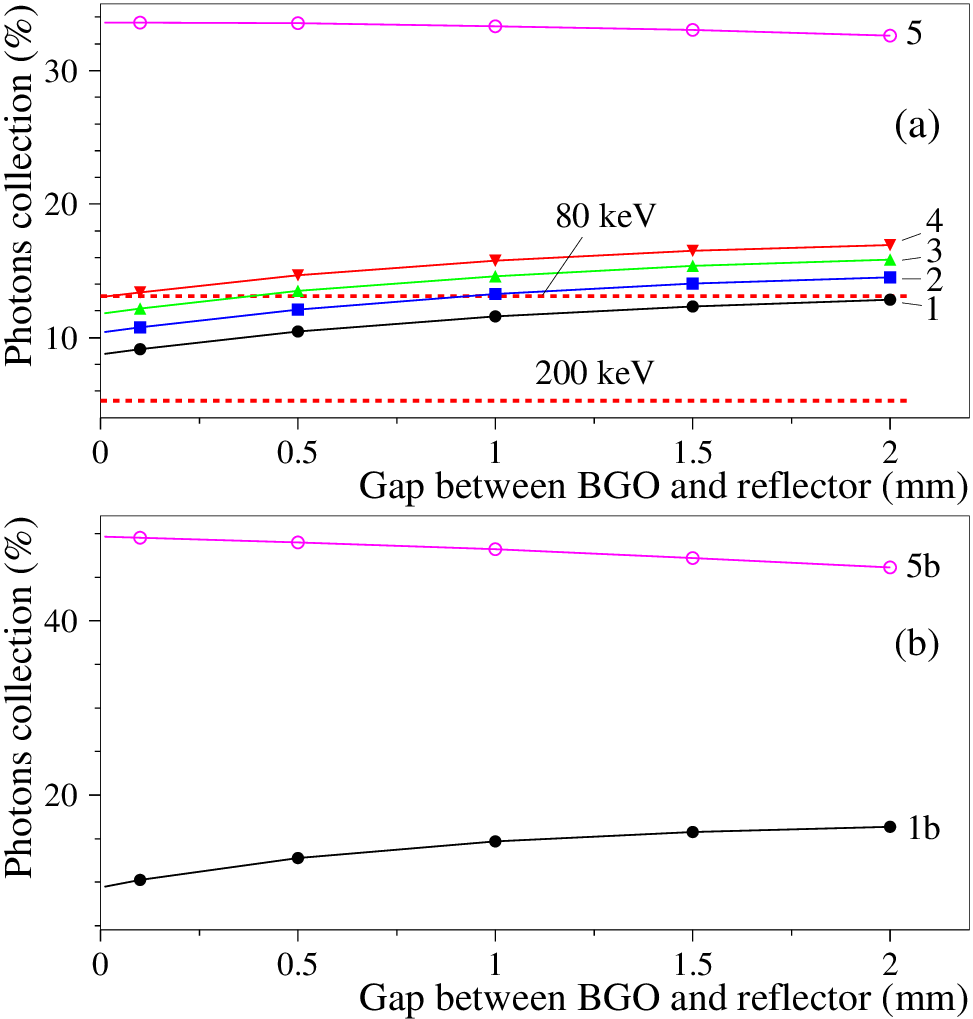,height=8.5cm}}
\caption{(a) Dependence of scintillation photon collection on the gap between the BGO scintillator and the reflector, as well as on the degree of GEANT4 specular reflectance (polishing) level set for all sides of the crystal, simulated for the BGO-veto prototype. The numbers ``1'' to ``4'' denote the polishing levels of the crystal from 100\% (``1'') to 97\% (``4'') with a 1\% step. The dashed red lines show the light-collection efficiency (5.25\% and 13.1\%), sufficient to meet the BGO module's 200 keV and 80 keV energy threshold, respectively, with full efficiency around 200 keV, as measured with the BGO-3 module (see Fig. \ref{fig:bgo-ly}). The points and line labelled as ``5'' are for a 100\% diffuse BGO crystal surfaces. (b) Simulated scintillation photon collection in the same geometry of the BGO-veto prototype, however, assuming the highly transparent BGO crystals (as in Fig. \ref{fig:emis-ref-transp} (c)). ``1b'' denotes 100\% polished scintillator, ``5b'' is for a 100\% diffuse BGO crystal.} 
 \label{fig:ProtoBGO-sim}
\end{figure}

In addition to the not very precisely known attenuation length of the material, several other parameters significantly affect the simulation results. E.g., the distance between the crystal and the reflector is not exactly fixed, since the reflective film can be deformed. In addition, only a limited number of the optical properties of elements like the ESR film are known. In particular, only a global reflectivity in the visible range of greater than 98\% is quoted by the producer. Variation of the reflection coefficient within 95\%--98\% leads to a decrease of the photon collection up to $6\%$. The degree of polishing of the crystal is most likely not 100\%, as some diffuse reflection can be observed. The effect of the crystal-surface quality and the distance between the crystal and the reflector is shown in Fig. \ref{fig:ProtoBGO-sim} (a). 
It can be seen that the diffusivity of the scintillation crystal surface improves light collection (which is in agreement with, e.g, \cite {Birks:1964,Danevich:2014a,Danevich:2014b}). Thus, one of the reasons for the poorer scintillation light collection in the prototype is the use of polished crystals.  
However, the main reason is the lower transparency of the BGO crystals used in the measurements. Simulations of light collection in the BGO-veto module geometry, assuming the high transparency (as in Fig. \ref{fig:emis-ref-transp} (c)) of the 100\% diffused BGO crystal, show that a 1.5-fold higher photon-collection efficiency can be achieved (see Fig. \ref{fig:ProtoBGO-sim} (b)). 

Regarding comparison of the light collection simulation with the results of the BGO prototype measurements, the achieved hardware threshold of the scintillation detector is $\sim80$ keV, corresponding to $5~RMS_{\rm{bsl}}$. This threshold is consistent with the optical simulations of the setup. However, pulses with higher amplitudes can drop out of the spectrum above this value, as shown in the energy spectrum in Fig. \ref{fig:bgo-ly}. It is due to pulse-shape cuts applied to reject spurious pulses and noise bursts, reducing the acceptance of physical events. A safe analysis threshold, with no distortion of the energy spectrum, can be set roughly at about $\sim 200$ keV.

Finally, it can be stated that our optical model, taking into account the inaccuracy of the BGO-module geometry and the optical properties of its elements, as well as a certain inaccuracy in determining the energy threshold in measurements with the prototype, satisfactorily describes the measurement results.

\section{Conclusions}
\label{conclusions}

Reducing the background is crucial for future experiments searching for neutrinoless double-beta decay, in which it is planned to increase the exposure to several tens of tonne $\times$ years. Achieving zero background will allow an experiment in which the sensitivity increases proportionally with the measurement time.

The BINGO project aims to develop new approaches to reducing the background in bolometric $0\nu2\beta$ experiments with $^{130}$Te and $^{100}$Mo to the level of $\sim 10^{-4}$ and $\sim 10^{-5}$ counts/(keV kg year), respectively. One of the steps to achieving such extremely low backgrounds is to use a cryogenic scintillation veto around the $2\beta$-decay-detector array.

As shown by Monte Carlo simulations of a $12\times12\times12$ detector array made of 4.5-cm-side cubic TeO$_2$ or Li$_2$MoO$_4$ crystals (951 kg and 487 kg, respectively), surrounded by a  5-cm thickness active BGO-crystal-scintillators veto with a 50 keV energy threshold, the background reduction from $^{208}$Tl in the low-temperature copper screen reaches factors of $\approx19$ and $\approx1.1\times10^3$ in the region of interest for $^{130}$Te ($2528\pm20$ keV) and $^{100}$Mo ($3034\pm50$ keV), respectively. The background reduction from $^{214}$Bi in the copper screen is $\approx54$ and $\approx11$ for TeO$_2$ and Li$_2$MoO$_4$ detectors, respectively. The advantage of an active screen is that even in the case of limited effectiveness of the active anti-coincidence veto (such as, for example, up to $^{214}$Bi in a copper low-temperature screens, when other $\gamma$ transitions do not accompany high-energy $\gamma$ quanta), the scintillation screen works as a ``spotless'' passive screen, thanks to a highly efficient veto of the background caused by its own radioactive contamination.

Nevertheless, this internal contamination of the veto is a point which will require careful consideration in future: in particular we plan to investigate the impact of potential failure or sub-optimal operation of individual detector(s) of the veto detector array and the resulting impact on expected backgrounds. Existing literature suggests that the resulting degradation may be rather minor: $^{238}$U and $^{232}$Th contaminations reported for BGO can be extremely low, with upper limits in \cite{Cardani:2012a} at the same level or better than the contamination of the internal Roman lead shield in the CUORE $0\nu2\beta$ experiment \cite{Adams:2024}. However, these limits are still a factor 4--7 higher than contamination of the Li$_2$MoO$_4$ crystals used in CUPID-Mo \cite{Augier:2013}, which warrants more careful investigations of both the contamination level and its possible impact. 

Potentially more problematic is the  presence of $^{207}$Bi at levels of $\sim 100$ mBq/kg in BGO \cite{Danevich:2018}, which combined with the slow response of cryogenic detectors may lead to a significant dead-time in anti-coincidence operation. A detailed study of this effect, evaluating the timing precision of NTL-assisted LD's of the veto together with anti-coincidence cuts leveraging both geometry and time information as used in CUORE \cite{Adams:2025b} is planned.  A new prototype with NTL-assisted LD's will be required to evaluate the time accuracy, which we plan to combine with more realistic simulations of the Mini-BINGO demonstrator.

Simulations of light propagation in a $100\times5\times5$ cm$^3$ scintillation counter based on BGO scintillation crystals have demonstrated the possibility of achieving a sufficiently low scintillation veto threshold of about $\sim50$ keV when using Ge cryogenic photodetectors. The highest light collection was achieved in a counter composed of two 50-cm-long scintillation elements ``face to face'', with 10\% diffusivity and fully optically diffused end faces. The use of photodetectors, which exploit the NTL effect for thermal signal amplification, allows further improvement of the scintillation counters energy threshold. 

Two prototypes of the veto system were developed and characterized, with significant progress achieved in the second version. This setup employed four trapezoidal BGO crystals, each read out by cryogenic light detectors, and was successfully operated in a cryogenic environment. Compared to the first prototype, the updated design ensured effective thermal decoupling of the slowly cooling BGO crystals from the sensitive bolometric elements. Light-to-heat ratio measurements at low temperatures were consistent with expectations for BGO, averaging 8.3 keV/MeV. 

Monte Carlo simulations of light photon collection in the BGO prototype were performed and showed a reasonable agreement with the measurement results where an energy threshold of $\sim$(80--200) keV was reached. The simulations also showed that a significant improvement in the threshold of the scintillation veto prototype can be achieved by using BGO crystals with higher optical transparency and a fully diffuse surface. 


\begin{acknowledgement}
This work is supported by the European Research Council (ERC) under the European Union’s Horizon 2020 research and innovation program (Project BINGO, Grant agreement ID: 865844). The cryostat used for the tests of the veto prototype here described — installed at ĲCLab (Orsay, France) — was donated
by the Dipartimento di Scienza e Alta Tecnologia of the Insubria University (Como, Italy). 

\end{acknowledgement}

\bibliographystyle{spphys}       
\bibliography{references}

@article{King:2014,
  title={Neutrino mass and mixing: from theory to experiment},
  author={King, Stephen F and Merle, Alexander and Morisi, Stefano and Shimizu, Yusuke and Tanimoto, Morimitsu},
  journal={New Journal of Physics},
  volume={16},
  number={4},
  pages={045018},
  year={2014},
  publisher={IOP Publishing}
}

@article{Bellini:2014,
  title={Neutrino oscillations},
  author={Bellini, G and Ludhova, L and Ranucci, Giorgia and Villante, FL},
  journal={Advances in High Energy Physics},
  volume={2014},
  number={1},
  pages={191960},
  year={2014},
  publisher={Wiley Online Library}
}

@article{Balantekin:2013,
  title={Neutrino oscillations},
  author={Balantekin, A B and Haxton, W C},
  journal={Progress in Particle and Nuclear Physics},
  volume={71},
  pages={150--161},
  year={2013},
  publisher={Elsevier}
}

@article{Bilenky:2015a,
  title={Neutrino in Standard Model and beyond},
  author={Bilenky, S M},
  journal={Physics of Particles and Nuclei},
  volume={46},
  number={4},
  pages={475--496},
  year={2015},
  publisher={Springer}
}

@article{Miranda:2015,
  title={Non standard neutrino interactions: current status and future prospects},
  author={Miranda, O G and Nunokawa, H},
  journal={New Journal of Physics},
  volume={17},
  number={9},
  pages={095002},
  year={2015},
  publisher={IOP Publishing}
}

@article{Xu:2021,
  title={Accurately weighing neutrinos with cosmological surveys},
  author={Xu, Weishuang Linda and DePorzio, Nicholas and Mu{\~n}oz, Julian B and Dvorkin, Cora},
  journal={Phys. Rev. D},
  volume={103},
  number={2},
  pages={023503},
  year={2021},
  publisher={APS}
}

@article{Shao:2025,
  title={Cosmological limits on the neutrino mass sum for beyond-$\Lambda$ CDM models},
  author={Shao, Helen and Givans, Jahmour J and Dunkley, Jo and Madhavacheril, Mathew and Qu, Frank J and Farren, Gerrit and Sherwin, Blake},
  journal={Phys. Rev. D},
  volume={111},
  number={8},
  pages={083535},
  year={2025},
  publisher={APS}
}

@article{Formaggio:2021,
  title={Direct measurements of neutrino mass},
  author={Formaggio, Joseph A and de Gouv{\^e}a, Andr{\'e} Luiz C and Robertson, RG Hamish},
  journal={Physics Reports},
  volume={914},
  pages={1--54},
  year={2021},
  publisher={Elsevier}
}

@article{Aker:2025,
  title={Direct neutrino-mass measurement based on 259 days of KATRIN data},
  author={Aker, M and others},
  journal={Science},
  volume={388},
  pages={6743},
  year={2025},
}

@article{Alpert:2025,
  title={Most Stringent Bound on Electron Neutrino Mass Obtained with a Scalable Low-Temperature Microcalorimeter Array},
  author={B K Alpert AND others},
  journal={Phys. Rev. Lett.},
  volume={135},
  pages={141801},
  year={2025},
}

@article{Bilenky:2015b,
  title={Neutrinoless double-beta decay: A probe of physics beyond the Standard Model},
  author={Bilenky, S M and Giunti, C},
  journal={International Journal of Modern Physics A},
  volume={30},
  number={},
  pages={1530001},
  year={2015},
  publisher={World Scientific}
}

@article{Vergados:2016,
  title={Neutrinoless double beta decay and neutrino mass},
  author={Vergados, Joannis D and Ejiri, Hiroyasu and {\v{S}}imkovic, Fedor},
  journal={International Journal of Modern Physics E},
  volume={25},
  number={11},
  pages={1630007},
  year={2016},
  publisher={World Scientific}
}

@article{Dolinski:2019,
  title={Neutrinoless double-beta decay: Status and prospects},
  author={Dolinski, Michelle J and Poon, Alan W P and Rodejohann, Werner},
  journal={Annual Review of Nuclear and Particle Science},
  volume={69},
  number={1},
  pages={219--251},
  year={2019},
  publisher={Annual Reviews}
}

@article{Gomez:2023,
  title={The search for neutrinoless double-beta decay},
  author={G{\'o}mez-Cadenas, Juan Jos{\'e} and Mart{\'\i}n-Albo, Justo and Men{\'e}ndez, Javier and Mezzetto, Mauro and Monrabal, Francesc and Sorel, Michel},
  journal={La Rivista del Nuovo Cimento},
  volume={46},
  number={10},
  pages={619--692},
  year={2023},
  publisher={Springer}
}

@article{Agostini:2023,
  title={Toward the discovery of matter creation with neutrinoless $\beta$ $\beta$ decay},
  author={Agostini, Matteo and Benato, Giovanni and Detwiler, Jason A and Men{\'e}ndez, Javier and Vissani, Francesco},
  journal={Reviews of Modern Physics},
  volume={95},
  number={2},
  pages={025002},
  year={2023},
  publisher={APS}
}

@article{Bossio:2024,
  title={Probing beyond the standard model physics with double-beta decays},
  author={Bossio, Elisabetta and Agostini, Matteo},
  journal={Journal of Physics G: Nuclear and Particle Physics},
  volume={51},
  number={2},
  pages={023001},
  year={2024},
  publisher={IOP Publishing}
}

@article{Acharya:2026,
  title={First Results on the Search for Lepton Number Violating Neutrinoless Double-$\beta$ Decay with the LEGEND-200 Experiment},
  XXXauthor={H.~Acharya AND N.~Ackermann AND M.~Agostini AND A.~Alexander AND C.~Andreoiu AND G.R.~Araujo AND F.T.~Avignone III AND M.~Babicz AND W.~Bae and others},
  author={H.~Acharya AND others},
  journal={Physical Review Letters},
  volume={136},
  pages={022701},
  year={2026},
  publisher={APS}
}

@article{Majorana:1937,
  title={Teoria simmetrica dell'elettrone e del positrone},
  author={E Majorana},
  journal={Il Nuovo Cimento},
  volume={14},
  pages={171–184},
  year={1937},
}

@article{Adams:2025a,
  title={Constraints on lepton number violation with the 2 tonne $\cdot$ year CUORE dataset},
  XXXauthor={Adams, D Q and Alduino, C and Alfonso, K and Avignone III, FT and Azzolini, O and Bari, G and Bellini, F and Benato, G and Biassoni, M and Branca, A and others},
  author={Adams, D Q and others},
  journal={Science},
  volume={390},
  pages={1029},
  year={2025},
}

@article{Azzolini:2022,
  title={Final Result on the Neutrinoless Double Beta Decay of $\^{}$\{$82$\}$ $ Se with CUPID-0},
  XXXauthor={Azzolini, O and Beeman, JW and Bellini, F and Beretta, M and Biassoni, M and Brofferio, C and Bucci, C and Capelli, S and Caracciolo, V and Cardani, L and others},
  author={Azzolini, O and others},
  journal={Phys. Rev. Lett.},
  volume={129},
  pages={111801},
year={2022}
}

@article{Augier:2022,
  title={Final results on the 0 $\nu$ $\beta$ $\beta$ decay half-life limit of 100 Mo from the CUPID-Mo experiment},
  XXXauthor={Augier, C and Barabash, AS and Bellini, F and Benato, G and Beretta, M and Berg{\'e}, L and Billard, J and Borovlev, Yu A and Cardani, L and Casali, N and others},
  author={Augier, C and others},
  journal={Eur. Phys. J. C},
  volume={82},
  number={11},
  pages={1033},
  year={2022},
  publisher={Springer}
}

@article{Abe:2025,
  title={Search for Majorana Neutrinos with the Complete KamLAND-Zen Dataset},
  XXXauthor={Abe, S and Araki, S and Chiba, K and Eizuka, M and Funahashi, Y and Furuto A and Gando, A and Gando and others},
  author={Abe, S and others},
  journal={Physical Review Letters},
  volume={135},
  pages={262501},
  year={2025},
  publisher={APS}
}

@article{Agrawal:2025a,
  title={Improved Limit on Neutrinoless Double Beta Decay of Mo 100 from AMoRE-I},
  XXXauthor={Agrawal, A and Alenkov, VV and Aryal, P and Beyer, J and Bhandari, B and Boiko, RS and Boonin, K and Buzanov, O and Byeon, CR and Chanthima, N and others},
  author={Agrawal, A and others},
  journal={Physical Review Letters},
  volume={134},
  number={8},
  pages={082501},
  year={2025},
  publisher={APS}
}

@article{Alfonso:2025,
  title={CUPID, the Cuore upgrade with particle identification},
  XXXauthor={Alfonso, K and Armatol, A and Augier, C and Avignone III, FT and Azzolini, O and Barabash, AS and Bari, G and Barresi, A and Baudin, D and Bellini, F and others},
  author={Alfonso, K and others},
  journal={Eur. Phys. J. C},
  volume={85},
  pages={737},
  year={2025},
  publisher={Springer}
}

@article{Alfonso:2026,
  title={Sensitivity of the CUPID experiment to $0\nu\beta\beta$ decay of $^{100}$Mo},
 author={Alfonso, K and others},
  journal={Eur. Phys. J. C},
  volume={86},
  pages={633},
  year={2025},
  publisher={Springer}
}

@article{Agrawal:2025b,
  title={Projected background and sensitivity of AMoRE-II},
  XXXauthor={Agrawal, A and Alenkov, VV and Aryal, P and Beyer, J and Bhandari, B and Boiko, RS and Boonin, K and Buzanov, O and Byeon, CR and Chanthima, N and others},
  author={Agrawal, A and others},
  journal={Eur. Phys. J. C},
  volume={85},
  number={1},
  pages={9},
  year={2025},
  publisher={Springer}
}

@article{Chernyak:2012,
  title={Random coincidence of 2 $\nu$ 2 $\beta$ decay events as a background source in bolometric 0 $\nu$ 2 $\beta$ decay experiments},
  author={Chernyak, DM and Danevich, FA and Giuliani, A and Olivieri, E and Tenconi, M and Tretyak, VI},
  journal={Eur. Phys. J. C},
  volume={72},
  number={4},
  pages={1989},
  year={2012},
  publisher={Springer}
}

@article{Christofferson:2018,
  title={Contamination control and assay results for the Majorana Demonstrator ultra clean components},
  XXXauthor={C D Christofferson AND  N Abgrall AND S I Alvis AND I J Arnquist AND F T Avignone III AND A S Barabash AND others},
  author={C D Christofferson AND others},
  journal={AIP Conference Proceedings},
  volume={1921},
  pages={060005},
  year={2018},
}

@misc{BINGO-web,
  title = {{BINGO experiment}},
  author = {},
  howpublished = {\url{http://www.bingo-neutrino.eu/}},
  note = {Accessed: 08 July 2025}}

@article{Khalife:2023,
  title={BINGO: Background reduction techniques for 0$\nu$2$\beta$ bolometric experiments},
  author={H Khalife AND others},
  journal={EPJ Web of Conferences},
  volume={290},
  pages={04002},
  year={2023},
}

@article{GEANT4,
  title={Geant4—a simulation toolkit},
XXXauthor={Agostinelli, Sea and Allison, John and Amako, K al and Apostolakis, John and Araujo, Henrique and Arce, Pedro and Asai, Makoto and Axen, D and Banerjee, Swagato and Barrand, GJNI and others},
author={Agostinelli, Sea and others},
  journal={Nucl. Instrum. Meth. Phys. Res. A},
  volume={506},
  number={3},
  pages={250--303},
  year={2003},
  publisher={Elsevier}
}

@article{DECAY0,
  title={Event generator DECAY4 for simulating double-beta processes and decays of radioactive nuclei},
  author={O. A.~Ponkratenko AND V. I.~Tretyak AND Yu. G.~Zdesenko},
  journal={Physics of Atomic Nuclei},
  volume={63},
  number={7},
  pages={1282--1287},
  year={2000},
  publisher={Springer}
}

@article{Sailer:2012,
  title={Low temperature light yield measurements in NaI and NaI (Tl)},
  author={Sailer, C and Lubsandorzhiev, B and Strandhagen, C and Jochum, J},
  journal={Eur. Phys. J. C},
  volume={72},
  number={6},
  pages={2061},
  year={2012},
  publisher={Springer}
}

@article{Mikhailik:2007a,
  title={Temperature dependence of CaMoO4 scintillation properties},
  author={Mikhailik, V B and Henry, S and Kraus, H and Solskii, I},
  journal={Nucl. Instrum. Meth. Phys. Res. A},
  volume={583},
  number={2-3},
  pages={350--355},
  year={2007},
  publisher={Elsevier}
}

@article{Moszynski:2003,
  title={Application of large area avalanche photodiodes to study scintillators at liquid nitrogen temperatures},
  author={Moszy{\'n}ski, M and Czarnacki, W and Klamra, Wlodzimierz and Szawlowski, M and Schotanus, P and Kapusta, M},
  journal={Nucl. Instrum. Meth. Phys. Res. A},
  volume={504},
  number={1-3},
  pages={307--312},
  year={2003},
  publisher={Elsevier}
}

@article{Moszynski:2004,
  title={Intrinsic energy resolution and light yield nonproportionality of BGO},
  author={Moszynski, Marek and Balcerzyk, Marcin and Czarnacki, W and Kapusta, Maciej and Klamra, Wlodzimierz and Syntfeld, A and Szawlowski, M},
  journal={IEEE Transactions on Nuclear Science},
  volume={51},
  number={3},
  pages={1074--1079},
  year={2004},
  publisher={IEEE}
}

@article{Gironnet:2008,
  title={Scintillation studies of Bi4Ge3O12 (BGO) down to a temperature of 6 K},
  author={Gironnet, J and Mikhailik, VB and Kraus, H and De Marcillac, P and Coron, N},
  journal={Nucl. Instrum. Meth. Phys. Res. A},
  volume={594},
  number={3},
  pages={358--361},
  year={2008},
  publisher={Elsevier}
}

@article{Moszynski:2005,
  title={Characterization of CaWO4 scintillator at room and liquid nitrogen temperatures},
 XXXauthor={Moszy{\'n}ski, M and Balcerzyk, M and Czarnacki, W and Nassalski, A and Szcz{\k{e}}{\'s}niak, T and Kraus, H and Mikhailik, V B and Solskii, I M},
  author={Moszy{\'n}ski, M and others},
  journal={Nucl. Instrum. Meth. Phys. Res. A},
  volume={553},
  number={3},
  pages={578--591},
  year={2005},
  publisher={Elsevier}
}

@article{Mikhailik:2007b,
  title={Scintillation studies of Ca WO 4 in the millikelvin temperature range},
  author={Mikhailik, V B and Kraus, H and Henry, S and Tolhurst, A J B},
  journal={Phys. Rev. B},
  volume={75},
  number={18},
  pages={184308},
  year={2007},
  publisher={APS}
}

@article{Pankratova:2024,
  title={Temperature behavior of Ce3+ emission in (Lu, Y) 2SiO5 single crystals excited by vacuum ultraviolet synchrotron light},
  author={Pankratova, Viktorija and Chernenko, K and Popov, Anatoli I and Pankratov, V},
  journal={Optical Materials: X},
  volume={22},
  pages={100322},
  year={2024},
  publisher={Elsevier}
}

@article{Kraus:2009,
  title={ZnWO4 scintillators for cryogenic dark matter experiments},
  XXXauthor={Kraus, H and Danevich, F A and Henry, S and Kobychev, V V and Mikhailik, V B and Mokina, V M and Nagorny, S S and Polischuk, O G and Tretyak, V I},
  author={Kraus, H and others},
  journal={Nucl. Instrum. Meth. Phys. Res. A},
  volume={600},
  number={3},
  pages={594--598},
  year={2009},
  publisher={Elsevier}
}

@article{Danevich:2010,
  title={Feasibility study of PbWO4 and PbMoO4 crystal scintillators for cryogenic rare events experiments},
  XXXauthor={Danevich, FA and Grinyov, BV and Henry, S and Kosmyna, MB and Kraus, H and Krutyak, N and Kudovbenko, VM and Mikhailik, VB and Nagornaya, LL and Nazarenko, BP and others},
  author={Danevich, F A and others},
  journal={Nucl. Instrum. Meth. Phys. Res. A},
  volume={622},
  number={3},
  pages={608--613},
  year={2010},
  publisher={Elsevier}
}

@article{Adams:2024,
  title={{Data-driven background model for the CUORE experiment}},
XXXauthor={Adams, DQ and Alduino, C and Alfonso, K and Avignone III, FT and Azzolini, O and Bari, G and Bellini, F and Benato, G and Beretta, M and Biassoni, M and others},
  author={Adams, D Q and others},
  journal={Phys. Rev. D},
  volume={110},
  number={5},
  pages={052003},
  year={2024},
  publisher={APS}
}

@article{Gusev:2001,
  title={Features of radiation damage of BGO crystals grown by the low-thermal-gradient Czochralski technique},
  author={V A Gusev AND others},
  journal={Nucl. Instrum. Meth. Phys. Res. A},
  volume={460},
  pages={457--464},
  year={2001},
  publisher={Elsevier}
}

@article{Ji:2014,
  title={Investigation of optical transmittance and light response uniformity of 600-mm-long BGO crystals},
  author={Ji, Zhiming and Ni, Haihong and Yuan, Lanying and Chen, Junfeng and Wang, Shaohua},
  journal={Nucl. Instrum. Meth. Phys. Res. A},
  volume={753},
  pages={143--148},
  year={2014},
  publisher={Elsevier}
}

@article{Holl:2002,
  title={A measurement of the light yield of common inorganic scintillators},
  author={Holl, Ina and Lorenz, Eckart and Mageras, Gikas},
  journal={IEEE Transactions on Nuclear Science},
  volume={35},
  number={1},
  pages={105--109},
  year={2002},
  publisher={IEEE}
}

@article{Humm:2003,
  title={From PET detectors to PET scanners},
  author={Humm, John L and Rosenfeld, Anatoly and Del Guerra, Alberto},
  journal={European Journal of Nuclear Medicine and Molecular Imaging},
  volume={30},
  number={11},
  pages={1574--1597},
  year={2003},
  publisher={Springer}
}

@article{Verdier:2011,
  title={Scintillation properties of Bi $\{$sub 4$\}$ Ge $\{$sub 3$\}$ O $\{$sub 12$\}$ down to 3 K under $\{$gamma$\}$ rays},
  author={M-A Verdier AND P Nadeau AND C Behan AND M Clavel AND C Dujardin},
  journal={Phys. Rev. B},
  volume={84},
  number={21},
  pages={214306},
  year={2011}
}

@article{Melcher:1985,
  title={Temperature dependence of fluorescence decay time and emission spectrum of bismuth germanate},
  author={Melcher, C L and Schweitzer, J S and Liberman, A and Simonetti, J},
  journal={IEEE Transactions on Nuclear Science},
  volume={32},
  number={1},
  pages={529--532},
  year={1985},
  publisher={IEEE}
}

@article{Weber:1973,
  title={Luminescence of Bi4 Ge3 O12: Spectral and decay properties},
  author={Weber, Marvin J and Monchamp, Roch R},
  journal={Journal of Applied Physics},
  volume={44},
  number={12},
  pages={5495--5499},
  year={1973},
  publisher={American Institute of Physics}
}

@article{Melcher:2000,
  title={Scintillation crystals for PET},
  author={Melcher, Charles L},
  journal={Journal of Nuclear Medicine},
  volume={41},
  number={6},
  pages={1051--1055},
  year={2000},
  publisher={Society of Nuclear Medicine}
}

@article{Jellison:2010,
  title={Optical properties of bismuth germanate},
  author={G E Jellison AND S Auluck AND D J Singh AND L A Boatner},
  journal={Journal of Applied Physics},
  volume={107},
  number={1},
  pages={013514},
  year={2010},
  publisher={AIP Publishing}
}

@article{Sasano:2013,
  title={Geometry dependence of the light collection efficiency of BGO crystal scintillators read out by avalanche photo diodes},
XXXauthor={Sasano, M and Nishioka, H and Okuyama, S and Nakazawa, K and Makishima, K and Yamada, S and Yuasa, T and Okumura, A and Kataoka, J and Fukazawa, Y and others},
   author={Sasano, M and others},
  journal={Nucl. Instrum. Meth. Phys. Res. A},
  volume={715},
  pages={105--111},
  year={2013},
  publisher={Elsevier}
}

@article{Brunner:2017,
  title={BGO as a hybrid scintillator/Cherenkov radiator for cost-effective time-of-flight PET},
  author={Brunner, S E and Schaart, D R},
  journal={Physics in Medicine \& Biology},
  volume={62},
  number={11},
  pages={4421},
  year={2017},
  publisher={IOP Publishing}
}

@article{Yuan:2018,
  title={High light response uniformity in industrial growth of 600-mm-long BGO crystals for dark matter particle explorer},
XXXauthor={Yuan, Lanying and Ni, Haihong and Ji, Zhiming and Chen, Junfeng and Song, Guilan and Qi, Xuejun and Li, Xiang and Sun, Shiyun and Wang, Shaohua},
author={Yuan, Lanying and others},
  journal={IEEE Transactions on Nuclear Science},
  volume={65},
  number={7},
  pages={1403--1407},
  year={2018},
  publisher={IEEE}
}

@article{Aspnes:1983,
  title={Dielectric functions and optical parameters of si, ge, gap, gaas, gasb, inp, inas, and insb from 1.5 to 6.0 ev},
  author={Aspnes, David E and Studna, A A},
  journal={Phys. Rev. B},
  volume={27},
  number={2},
  pages={985},
  year={1983},
  publisher={APS}
}

@article{Nunley:2016,
  title={Optical constants of germanium and thermally grown germanium dioxide from 0.5 to 6.6 eV via a multisample ellipsometry investigation},
XXXauthor={Nunley, Timothy Nathan and Fernando, Nalin S and Samarasingha, Nuwanjula and Moya, Jaime M and Nelson, Cayla M and Medina, Amber A and Zollner, Stefan},
author={Nunley, Timothy Nathan and others},
  journal={Journal of Vacuum Science \& Technology B, Nanotechnology and Microelectronics: Materials, Processing, Measurement, and Phenomena},
  volume={34},
  number={6},
  pages={061205},
  year={2016},
  publisher={American Vacuum Society}
}

@article{Alfonso:2023,
  title={A first test of CUPID prototypal light detectors with NTD-Ge sensors in a pulse-tube cryostat},
XXXauthor={Alfonso, K and Armatol, A and Augier, C and Avignone, FT and Azzolini, O and Balata, M and Barabash, AS and Bari, G and Barresi, A and Baudin, D and others},
author={Alfonso, K and others},
  journal={Journal of Instrumentation},
  volume={18},
  number={06},
  pages={P06033},
  year={2023},
  publisher={IOP Publishing}
}

@article{Novati:2019,
  title={Charge-to-heat transducers exploiting the Neganov-Trofimov-Luke effect for light detection in rare-event searches},
XXXauthor={Novati, V and Berg{\'e}, L and Dumoulin, L and Giuliani, A and Mancuso, M and De Marcillac, P and Marnieros, S and Olivieri, E and Poda, DV and Tenconi, M and others},
  author={Novati, V and others},
  journal={Nucl. Instrum. Meth. Phys. Res. A},
  volume={940},
  pages={320--327},
  year={2019},
  publisher={Elsevier}
}

@article{Armatol:2024,
  title={BINGO innovative assembly for background reduction in bolometric 0$\nu$$\beta$$\beta$ experiments},
XXXauthor={Armatol, A and Augier, C and Bandac, IC and Baudin, D and Benato, G and Berest, V and Berg{\'e}, L and Billard, J and Calvo-Mozota, JM and Carniti, P and others},
  author={Armatol, A and others},
  journal={Nucl. Instrum. Meth. Phys. Res. A},
  volume={1069},
  pages={169936},
  year={2024},
  publisher={Elsevier}
}

@misc{tekraMelinexxAEFilm,
  title = {Melinex 339 Film},
  author = {},
  howpublished = {\url{https://www.tekra.com/products/films/polyester-films/polyester-pet/melinex-339}},
  note = {Accessed: 08 July 2025}}

@article{Haller:1994,
  title={Advanced far-infrared detectors},
  author={Haller, E E},
  journal={Infrared Physics \& Technology},
  volume={35},
  number={2-3},
  pages={127--146},
  year={1994},
  publisher={Elsevier}
}

@article{heater,
  title={Methods for response stabilization in bolometers for rare decays},
XXXauthor={Alessandrello, A and Brofferio, C and Bucci, C and Cremonesi, O and Giuliani, A and Margesin, Benno and Nucciotti, A and Pavan, M and Pessina, G and Previtali, E and others},
  author={Alessandrello, A and others},
  journal={Nucl. Instrum. Meth. Phys. Res. A},
  volume={412},
  number={2-3},
  pages={454--464},
  year={1998},
  publisher={Elsevier}
}

@article{Poda:2021hsv,
  title={Scintillation in low-temperature particle detectors},
  author={Poda, Denys},
  journal={Physics},
  volume={3},
  number={3},
  pages={473--535},
  year={2021},
  publisher={MDPI}
}

@book{Birks:1964,
  title={The Theory and Practice of Scintillation Counting},
  author={Birks, J B},
  publisher="{Pergamon Press, Oxford}",
  year={1964}}

@article{Danevich:2014a,
  title={Optimization of light collection from crystal scintillators for cryogenic experiments},
  author={Danevich, F A and Kobychev, R V and Kobychev, V V and Kraus, H and Mikhailik, V B and Mokina, V M},
  journal={Nucl. Instrum. Meth. Phys. Res. A},
  volume={744},
  pages={41--47},
  year={2014},
  publisher={Elsevier}
}

@article{Danevich:2014b,
  title={Impact of geometry on light collection efficiency of scintillation detectors for cryogenic rare event searches},
XXXauthor={Danevich, FA and Kobychev, VV and Kobychev, RV and Kraus, H and Mikhailik, VB and Mokina, VM and Solsky, IM},
author={Danevich, F A and others},
  journal={Nucl. Instrum. Meth. Phys. Res. B},
  volume={336},
  pages={26--30},
  year={2014},
  publisher={Elsevier}
}

@article{CUPID:2023dqd,
  title={Twelve-crystal prototype of Li2MoO4 scintillating bolometers for CUPID and CROSS experiments},
XXXauthor={Alfonso, K and Armatol, A and Augier, C and Avignone, FT and Azzolini, O and Balata, M and Bandac, IC and Barabash, AS and Bari, G and Barresi, A and others},
author={Alfonso, K and others},
  journal={Journal of Instrumentation},
  volume={18},
  number={06},
  pages={P06018},
  year={2023},
  publisher={IOP Publishing}
}

@article{CROSS_Run9:2026,
  author = "Armatol, A and others",
  title = "{Cryogenic light detectors with thermal signal amplification for $0\nu\beta\beta$ search experiments}",
  journal = "Journal of Instrumentation",
  volume = "21",
  pages = "P01035",
  year = "2026",
  eprint         = "2507.15732",
  archivePrefix  = "arXiv",
  primaryClass   = "physics.ins-det"
}

@article{Mancuso:2014a,
  author = "Mancuso, M and others",
  title = "{An aboveground pulse-tube-based bolometric test facility for the validation of the LUMINEU ZnMoO$_4$ crystals}",
  journal = "J. Low Temp. Phys.",
  volume = "176",
  pages = "571",
  year = "2014"
}

@article{Cardani:2012a,
  author = "L Cardani AND S Di Domizio AND L Gironi",
  title = "{A BGO scintillating bolometer for $\gamma$ and $\alpha$ spectroscopy}",
  journal = "Journal of Instrumentation",
  volume = "7",
  pages = "P10022",
  year = "2012"
}

@article{Augier:2013,
  author = "C Augier AND others",
  title = "{The background model of the CUPID-Mo $0\nu\beta\beta$ experiment}",
  journal = "Eur. Phys. J. C",
  volume = "83",
  pages = "675",
  year = "2023"
}

@article{Danevich:2018,
  author = "F A Danevich AND V I Tretyak",
  title = "{Radioactive contamination of scintillators}",
  journal = "International Journal of Modern Physics A",
  volume = "33",
  pages = "1843007",
  year = "2018"
}

@misc{Adams:2025b,
      title={{End-to-End Data Analysis Methods for the CUORE Experiment}}, 
      author={D. Q. Adams AND others},
      year={2025},
      eprint={2510.25720},
      archivePrefix={arXiv},
      primaryClass={nucl-ex},
      url={https://arxiv.org/abs/2510.25720}, 
}

@article{ANDREOTTI2012161,
title = {Production, characterization, and selection of the heating elements for the response stabilization of the CUORE bolometers},
journal = {Nucl. Instrum. Meth. Phys. Res. A},
volume = {664},
number = {1},
pages = {161-170},
year = {2012},
issn = {0168-9002},
XXXdoi = {https://doi.org/10.1016/j.nima.2011.10.065},
XXurl = {https://www.sciencedirect.com/science/article/pii/S0168900211020316},
XXXauthor = {Erica Andreotti and Chiara Brofferio and Luca Foggetta and Andrea Giuliani and Benno Margesin and Claudia Nones and Marisa Pedretti and Claudia Rusconi and Chiara Salvioni and Margherita Tenconi},
author = {Erica Andreotti and others},
XXkeywords = {Detectors of radiation, Bolometers, Stabilization of response},
XXabstract = {One of the critical issues while operating bolometric detectors over periods of time of 1 year or more consists of keeping their response stable within a 0.1% level, despite the unavoidable temperature fluctuations of the cryogenic set-up. By using an energy pulser, which periodically delivers a fixed amount of energy in the absorber, it is possible to stabilize the response of the bolometers. A stabilization technique using heating devices, made up of heavily doped semiconductor material (well above the metal-to-insulator transition), has been developed in the framework of the CUORE experiment. In this paper we describe in detail the procedure for the realization of the heating elements, based on silicon planar technology. We then report on the multi-step low temperature characterization (77K, 4.2K, 1.5K, 35mK) of the heaters. Finally, an example of achieved stabilization for a CUORE-like detector is reported. The ∼1500 heaters tested at ∼1.5K show less than 0.5% change in resistance between 30μV and 3mV, and less than 1% change in value between 50mK and 800mK. In particular, the resistance change between 4.2K and 1.5K is less than 0.1%.}
}

@phdthesis{Armatol:PhD,
  author       = {A Armatol}, 
  title        = {Innovative methods for background rejection in next-generation
neutrinoless double beta decay bolometric experiments},
  school       = {Universit\'{e} aris-Saclay, Gif-sur-Yvette, France},
  year         = 2023,
}

@article{Mancuso:2014,
  title={An experimental study of antireflective coatings in Ge light detectors for scintillating bolometers},
  author={M Mancuso AND others},
  journal={EPJ Web of Conferences},
  volume={65},
  pages={04003},
  year={2018},
}

@article{Mcgregor:2018scintillators,
  title={Materials for gamma-ray spectrometers: Inorganic scintillators},
  author={McGregor, Douglas S},
  journal={Annual Review of Materials Research},
  volume={48},
  pages={245--277},
  year={2018}
}

@article{Armengaud:2017,
  author = "Armengaud, E and others",
  title = "{Development of $^{100}$Mo-containing scintillating bolometers for a high-sensitivity neutrinoless double-beta decay search}",
  journal = "Eur. Phys. J. C",
  volume = "77",
  pages = "785",
  year = "2017"}

@article{Giuliani:2026,
  author = {Giuliani, A.},
  title = {{The Quest for Neutrinoless Double Beta Decay: Progress and Prospects}},
  journal = {Physics and the Cosmos},
  volume = {1},
  pages = {6},
  year = {2026}}

\end{document}